\documentclass[%
reprint,
superscriptaddress,
amsmath,amssymb,
aps,
pra,
]{revtex4-2}
\usepackage{color}
\usepackage{soul}
\usepackage{rotating}
\usepackage{amsmath}
\usepackage{mathrsfs}
\usepackage{graphicx}% Include figure files
\usepackage{dcolumn}% Align table columns on decimal point
\usepackage{bm}% bold math
\usepackage{physics}     %物理公式
\usepackage{hyperref}% add hypertext capabilities
\usepackage{subfigure}
\usepackage{hyperref}% add hypertext capabilities
\usepackage{algorithmicx}
\usepackage{algpseudocode}

\hypersetup{
	colorlinks = true, %将链接文字带颜色
	linkcolor = blue, %图表引用颜色设置为蓝色
	urlcolor = blue,%网页链接为蓝色
	citecolor = blue,%文献引用颜色设置为蓝色
}

\begin{document}
\title{Covertness as a resource constraint in quantum target sensing}

\author{Yu Yang}
\email{yangyu1229@hotmail.com} 
\affiliation{Ministry of Education Key Laboratory for Nonequilibrium Synthesis and Modulation of Condensed Matter, Shaanxi Province Key Laboratory of Quantum Information and Quantum Optoelectronic Devices, School of Physics, Xi’an Jiaotong University, Xi’an 710049, China}

\author{Fengyu Guan}
\affiliation{Ministry of Education Key Laboratory for Nonequilibrium Synthesis and Modulation of Condensed Matter, Shaanxi Province Key Laboratory of Quantum Information and Quantum Optoelectronic Devices, School of Physics, Xi’an Jiaotong University, Xi’an 710049, China}

\author{Shiping Guo}
\affiliation{Ministry of Education Key Laboratory for Nonequilibrium Synthesis and Modulation of Condensed Matter, Shaanxi Province Key Laboratory of Quantum Information and Quantum Optoelectronic Devices, School of Physics, Xi’an Jiaotong University, Xi’an 710049, China}

\author{Ruifeng Liu}
\affiliation{Ministry of Education Key Laboratory for Nonequilibrium Synthesis and Modulation of Condensed Matter, Shaanxi Province Key Laboratory of Quantum Information and Quantum Optoelectronic Devices, School of Physics, Xi’an Jiaotong University, Xi’an 710049, China}

\author{Pei Zhang}
\affiliation{Ministry of Education Key Laboratory for Nonequilibrium Synthesis and Modulation of Condensed Matter, Shaanxi Province Key Laboratory of Quantum Information and Quantum Optoelectronic Devices, School of Physics, Xi’an Jiaotong University, Xi’an 710049, China}

\author{Fuli Li}
\affiliation{Ministry of Education Key Laboratory for Nonequilibrium Synthesis and Modulation of Condensed Matter, Shaanxi Province Key Laboratory of Quantum Information and Quantum Optoelectronic Devices, School of Physics, Xi’an Jiaotong University, Xi’an 710049, China}

\date{\today}

\begin{abstract}
Quantum covert target sensing aims to detect low-reflectivity targets embedded in the thermal environment while avoiding discovery by an adversary.
This covertness requirement constrains the admissible probe energy, thereby imposing a trade-off between covertness and sensing performance.
However, how this resource constraint determines the ultimate estimation precision of the target parameter remains largely unexplored.
To fill this gap, we establish a unified framework connecting covert target discrimination with quantum-enhanced parameter estimation.
We first present perturbative $\epsilon$-covertness-induced bounds on the probe energy, which yield an admissible energy window narrowing as $\mathcal{O}(\sqrt{\epsilon/M})$.
Moreover, we introduce the covertness-constrained quantum Fisher information (CCQFI) to quantify the ultimate precision of target-reflectivity estimation and assess its attainability with specific measurements.
For Gaussian-distributed coherent-state (GCS) and two-mode squeezed vacuum (TMSV) probes, their CCQFIs scale as $\mathcal{O}(\epsilon)$ and $\mathcal{O}(M)+\mathcal{O}(\sqrt{\epsilon M})+\mathcal{O}(\epsilon)$, respectively.
In particular, in the strong covertness regime $\epsilon\to0$, the GCS CCQFI vanishes, whereas the TMSV retains a finite $\mathcal{O}(M)$ contribution exhibiting a pronounced metrological advantage.
Besides, photon counting saturates the GCS CCQFI, whereas homodyne detection achieves a higher attainability for the TMSV probe.
Our work identifies covertness as a quantitative resource constraint governing the metrological limits of quantum target sensing.
\end{abstract}

\maketitle

\section{Introduction}\label{Sec:Introduction}
Quantum illumination~\cite{QI1,QI2,QI3,QI4,QI5,QI6,QI7} aims to detect a target embedded in a noisy environment by leveraging quantum probes, exhibiting an enhanced error exponent over conventional probes.
Incorporating security and covertness~\cite{SecureCom,SecureChannel} into this scenario motivates quantum covert target detection~\cite{E2022,Gu2024,PRL2026}.
Specifically, the transmitter (Alice) sends a quantum probe toward a low-reflectivity target, while the receiver (Charlie) collects the returned field to discriminate between the target-absent and target-present hypotheses.
Meanwhile, an adversary (Eve) monitors the surrounding environment to determine whether Alice is probing the target. 
Increasing the probe energy generally improves Charlie's target discrimination but makes Alice's probing activity more detectable to Eve in parallel.
Consequently, covertness constrains the admissible probe energy, thereby rendering the probe brightness a constrained resource.
Recent studies have demonstrated entanglement-enhanced covert target discrimination~\cite{E2022,Gu2024,Gaussian_Review}, and established fundamental bounds on the discrimination error, such as the quantum Chernoff bound (QCB)~\cite{QCB1,QCB2,PRL2026}. Binary hypothesis testing has also been extended to multiple-hypothesis discrimination, thereby enabling quantum ranging~\cite{Range1,Range2,Range}.

However, these studies have focused primarily on discrimination errors within the framework of quantum hypothesis testing.
Beyond binary detection, a fundamental question is how covertness constrains the sensing resources available for parameter estimation and, consequently, the ultimate metrological precision. 
For single parameter estimation, such as the target's reflectivity, the ultimate precision is generally bounded by the quantum Fisher information (QFI)~\cite{Paris2009,Review1,Liu2020}. 
Although relative entropy~\cite{E2022} and necessary covertness condition~\cite{Gu2024} have been proposed to characterize adversarial distinguishability and numerically bound the probe energy, an explicit analytical relation between covertness and the admissible probe-energy window remains lacking. 
Consequently, how the resulting covertness-induced resource constraint reshapes the QFI and its asymptotic scaling remains largely unexplored. 
Moreover, continuous-variable Gaussian probes~\cite{Review1,Review2}, especially the two-mode squeezed-vacuum (TMSV) state, have demonstrated advantages in target discrimination~\cite{QI1,ES} and quantum ranging~\cite{Range1,Range2}. 
Whether such quantum advantages survive in parameter estimation under covertness-constrained sensing resources remains an open question.

In this work, we establish a unified framework that connects covert target discrimination with quantum-enhanced parameter estimation under covertness-constrained sensing resources.
For two canonical probes, the Gaussian-distributed coherent state (GCS) and the TMSV state, we first investigate their discrimination performance through quantum Chernoff bounds and ascertain the entanglement advantage at fixed signal brightness.
We then establish and analytically prove perturbative $\epsilon$-covertness-induced bounds on the probe energy, which confine the admissible probe brightness to a finite window around the thermal-background level.
Under this resource constraint, the covertness-constrained quantum Fisher information (CCQFI) is introduced to quantify the metrological precision limit for target-reflectivity estimation. 
Finally, we assess the attainability of the CCQFI using the covertness-constrained classical Fisher information (CFI) associated with two canonical measurements, homodyne and photon-counting measurements.

The remainder of this paper is organized as follows.
Sec.~\ref{Sec:model} introduces the model of quantum covert sensing.
In Sec.~\ref{Sec:DS}, we study target discrimination under the covertness constraint using quantum Chernoff bounds.
The perturbative covertness-induced bounds on the probe energy are developed in Sec.~\ref{Sec:bound}.
Sec.~\ref{Sec:CCQFI} proposes the CCQFI and evaluates it for different probes.
Sec.~\ref{Sec:Meas} examines its attainability under homodyne and photon-counting measurements.
Sec.~\ref{Sec:conclusion} finally summarizes this paper.

\section{Model}\label{Sec:model}
\begin{figure}[htp]
	\centering
	\includegraphics[width=0.45\textwidth]{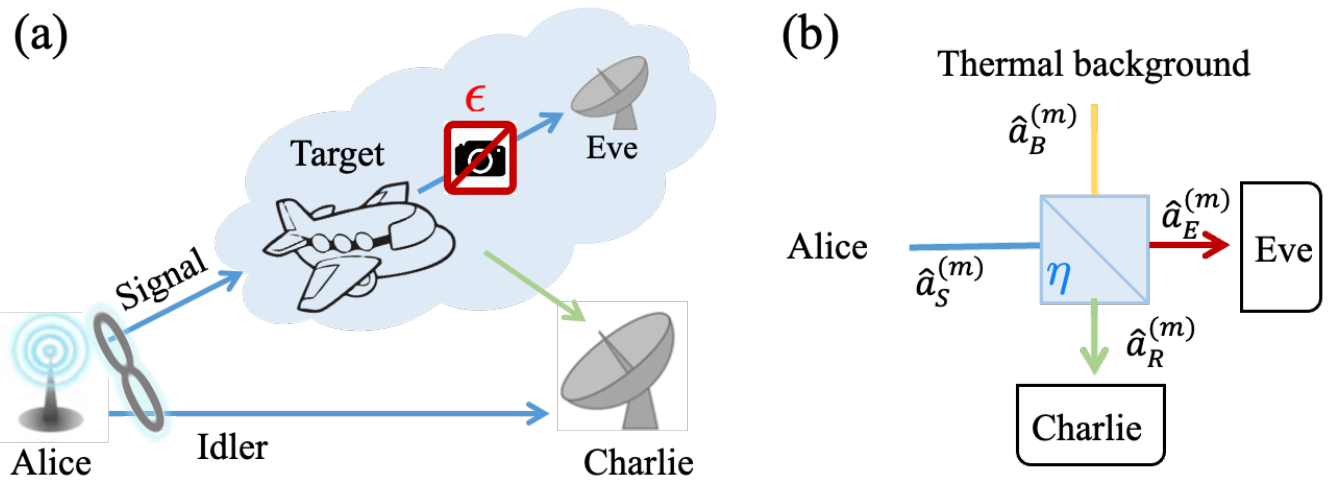}
	\caption
	{(a) Schematic diagram of quantum covert target detection with the covertness $\epsilon$.
    (b) Field-operator relations for a target modeled as a beam splitter with reflectivity $\eta$.
	Here $a_S^{(m)}$, $a_B^{(m)}$, $a_R^{(m)}$ and $a_E^{(m)}$ denote the annihilation operators of the $m$-th signal, thermal-background, return, and environmental modes with $m=1,2,\cdots M$, respectively.}
	\label{Fig.model}
\end{figure}

The schematic diagram for quantum covert target detection is shown in Fig.~\ref{Fig.model} (a).
The transmitter (Alice) prepares $M$ independent and identically distributed (i.i.d.) signal-idler probe states. 
For each copy, the signal mode is transmitted toward a low-reflectivity target, while the corresponding idler mode is retained and made available to the receiver (Charlie) for joint processing with the returned mode.
Charlie collects the returned modes together with the retained idler modes.  
The adversary (Eve) monitors the environmental output to determine whether Alice is probing the target. 
This probing strategy is required to satisfy the $\epsilon$-covertness condition
\begin{eqnarray}
P_e^E \geq \frac{1}{2}-\epsilon,
\end{eqnarray}
where $P_e^E$ denotes Eve's minimum error probability in discriminating between the absence and presence of Alice's probing activity.
The parameter $\epsilon$ quantifies the deviation from perfect covertness, where $\epsilon=0$ yields the random-guessing limit $P_e^E=1/2$.
The target is modeled as a beam splitter with low-reflectivity $\eta$, as illustrated in Fig.~\ref{Fig.model} (b). 
Here $\eta=0$ represents the absence of the target, while $\eta>0$ corresponds to its presence.

The thermal-background state in the Fock basis $|n\rangle$ is written into
\begin{eqnarray}\label{eq:back}
&& \rho_{\rm th}=\bigotimes_{m=1}^M \rho_{\rm th}^{(m)}(N_B), \nonumber\\
&& \rho_{\rm th}^{(m)}(N_B)=\sum_{n=0}^{\infty}\frac{N_B^n}{(N_B+1)^{n+1}}|n\rangle\langle n|,
\end{eqnarray}
where $N_B$ represents the mean background photon number per mode.
Accordingly, the total mean background photon number over the $M$ modes is $M N_B$.
We next derive the corresponding hypothesis states observed by Charlie and Eve, respectively.

\paragraph{Receiver (Charlie).}
Alice prepares a signal-idler probe state $\rho_{SI}$, where the signal mode is transmitted toward the target and the idler mode, if present, is retained locally.
Assuming equal prior probabilities for the absence and presence of the target, Charlie's corresponding M-copy hypothesis states are 
\begin{eqnarray}\label{eq:Cstate}
\rho_{0,{C}}&=&\operatorname{Tr}_{S}[\rho_{SI}] \otimes \rho_\text{th},\nonumber\\
\rho_{1,{C}}&=& \operatorname{Tr}_{E} \left[ U_{BS} \left(\rho_{SI} \otimes \rho_\text{th} \right)  U_{BS}^\dagger  \right],
\end{eqnarray}  
where $\operatorname{Tr}_S[\bullet]$ denotes the partial trace over the signal part in $\rho_{SI}$, and $\operatorname{Tr}_E[\bullet]$ denotes the partial trace over the environmental output modes. 
Here, $U_{BS}=\bigotimes_{m=1}^M U_{BS}^{(m)}$ represents the beam-splitter transformation induced by the target, which acts on the signal and background modes while leaving the idler modes unchanged.
The $m$-th return mode is formulated as
\begin{eqnarray}
a_R^{(m)}=\sqrt{\eta} a_S^{(m)}+\sqrt{1-\eta} a_B^{(m)}\;,
\end{eqnarray} 
for $m=1,2,\cdots M$, 
where $a_S^{(m)}$, $a_B^{(m)}$ and $a_R^{(m)}$ denote the annihilation operators of the transmitted signal, thermal-background, and received modes, respectively.

\paragraph{Adversary (Eve).}
Eve monitors the environmental output that may contain the probe signal leaked by Alice.
Its M-copy hypothesis states in the absence and presence of Alice's probing behavior are 
\begin{eqnarray}\label{eq:rhoE}
\rho_{0,E}&=&\rho_{\rm th},\nonumber\\
\rho_{1,E}&=&\operatorname{Tr}_{R}\left[U_{\rm BS}\left(\operatorname{Tr}_{I}[\rho_{SI}] \otimes\rho_{\rm th}\right)U_{\rm BS}^{\dagger}\right],
\end{eqnarray}
where $\operatorname{Tr}_I[\bullet]$ denotes the partial trace over the idler part in $\rho_{SI}$, and $\operatorname{Tr}_R[\bullet]$ denotes the partial trace over the return mode at Charlie's side.
The $m$-th environmental output mode obeys
\begin{eqnarray}\label{eq:rela}
a_E^{(m)}=\sqrt{1-\eta}\,a_S^{(m)}-\sqrt{\eta}\,a_B^{(m)}.
\end{eqnarray}

In this paper, we mainly focus on two representative continuous-variable Gaussian probes.
The first is a GCS probe, whose ensemble-averaged density operator is
\begin{eqnarray}\label{eq:GCS}
	\rho_{\rm GCS}= \bigotimes_{m=1}^{M} \int d^2\alpha_m\, P(\alpha_m) |\alpha_m\rangle\langle\alpha_m|, 
\end{eqnarray}
with
\begin{eqnarray}
	P(\alpha_m)=\frac{1}{\pi N_s} e^{-\frac{|\alpha_m|^2}{N_s}},
\end{eqnarray}
where $N_s=\int d^2\alpha P(\alpha) |\alpha|^2$ denotes the mean transmitted signal photon number per mode.
Here, $\alpha_m=\alpha_{m,r}+i \alpha_{m,i}$ is the complex coherent-state amplitude of the $m$-th mode.
Eq.~(\ref{eq:GCS}) corresponds to the Glauber-Sudarshan P-representation~\cite{QOptics} of a thermal state that mimics the statistics of the thermal background.
Particularly, we assume that Charlie does not know the specific realization of the coherent-state amplitude $\alpha_m$ in each mode emitted by Alice, but only its probability distribution. 
Hence, no auxiliary classical channel is required to convey the amplitudes $\{\alpha_m\}$ to Charlie. 
This differs from the covert sensing protocol in Ref.~\cite{Gu2024}, where the coherent-state amplitudes are shared through an additional classical channel.

The second probe is the two-mode squeezed-vacuum (TMSV) state
\begin{eqnarray}\label{eq:TMSV}
	\rho_{\rm TMSV}=\left( |\psi_{\rm TMSV}\rangle \langle\psi_{\rm TMSV}|\right)^{\otimes M}, 
\end{eqnarray}
where each signal-idler pair is
\begin{eqnarray}\label{eq:TMSVsingle}
	|\psi_{\rm TMSV}\rangle= \sum_{n=0}^{\infty} \sqrt{\frac{N_s^n}{(N_s+1)^{n+1}}} |n\rangle_S|n\rangle_I.
\end{eqnarray}
The signal modes are transmitted toward the target, whereas the corresponding idler modes are stored locally for the subsequent joint measurements with the returned modes.
Tracing out the idler of each TMSV pair yields
$\operatorname{Tr}_{I}[\rho_\text{TMSV}]=\bigotimes_{m=1}^M \rho_{\rm th}^{(m)}(N_s)$, 
which is identical to the ensemble-averaged GCS state with the same mean signal photon number $N_s$.
Thus, the GCS and TMSV probes exhibit identical states to Eve for a fixed $N_s$.

We next investigate two complementary sensing tasks under the covertness constraint.
One is binary target discrimination, and another is target-reflectivity estimation.
The former is evaluated by the discrimination error probability, while the latter is quantified by the covertness-constrained quantum Fisher information. 
This allows us to connect the hypothesis-testing and metrological precision of covert target sensing within a unified framework.

\section{Discrimination error probability under the covertness constraint}\label{Sec:DS}
\subsection{Lower bound and quantum Chernoff bound}
The covertness requirement brings a lower bound on the discrimination error probability. 
Specifically, for Charlie, the target detection constitutes a binary hypothesis-testing problem between the target-absent state $\rho_{0,C}$ and the target-present state $\rho_{1,C}$. 
For equal prior probabilities, optimization over all positive-operator-valued measurements (POVMs) gives the minimum discrimination error probability, i.e., the Helstrom bound~\cite{QCB1,Helstrom} $P_e^C$, which satisfies
\begin{eqnarray}
L_e^C(\epsilon) \le P_e^C=\frac{1}{2}\left(1-\frac{|| \rho_{0,C}-\rho_{1,C} ||_1}{2} \right) \le  T_e^C.
\end{eqnarray}
Here, $||A||_1=\operatorname{Tr}|A|=\operatorname{Tr}\sqrt{A ^\dagger A}$ denotes the trace norm of operator $A$.
The covertness constraint leads to a probe-independent lower bound~\cite{Gu2024}
\begin{eqnarray}\label{eq:Lc}
L_e^C(\epsilon)=\frac{1}{2} \left({1-\sqrt{1-(1-2\epsilon)^4 f^{2M}}}\right),
\end{eqnarray}
for $f=\frac{\left( N_B+1-\frac{N_B}{x}\right)[\eta N_B(1-x)+1]}{\sqrt{(N_B+1)[(1-\eta)N_B+1]}-N_B \sqrt{ 1-\eta}},
x=1-\frac{1-\Theta}{(1-\eta)-\eta N_B(1-\Theta)}$ and $\Theta=\sqrt{\frac{(1-\eta)(N_B+1)}{1+(1-\eta)N_B}}$.
For a given probe state, an upper bound on the minimum discrimination error is provided by the quantum Chernoff bound~\cite{QCB,QCB1}
\begin{eqnarray}
T_e^C=\frac{1}{2}  \inf_{0 \le s \le 1} \operatorname{Tr}[\rho_{0,C}^s \rho_{1,C}^{1-s}],
\end{eqnarray}
where the infimum is taken over $s\in[0,1]$.

By contrast, Eve performs a binary discrimination between the no-probing state $\rho_{0,E}$ and the probing-active state $\rho_{1,E}$.
For equal priors, the minimum error probability $P_e^E$ is lower bounded by the covertness threshold and upper bounded by the corresponding QCB, i.e.,
\begin{eqnarray}\label{eq:Eveerror}
\frac{1}{2}-\epsilon \le P_e^E=\frac{1}{2}\left(1-\frac{|| \rho_{0,E}-\rho_{1,E} ||_1}{2} \right) \le  T_e^E,
\end{eqnarray}
with
\begin{eqnarray}
T_e^E=\frac{1}{2}  \inf_{0 \le s \le 1} \operatorname{Tr}[\rho_{0,E}^s \rho_{1,E}^{1-s}].
\end{eqnarray}

\subsection{Entanglement-enhanced discrimination and detectability}
Both the GCS of Eq.~(\ref{eq:GCS}) and the TMSV of Eq.~(\ref{eq:TMSV}) probes yield Gaussian hypothesis states through the thermal-loss channel.
Particularly, we assume that Charlie has no access to Alice's random coherent amplitudes, thereby avoiding an auxiliary classical channel and associated information leakage.
Using the mean vectors and Wigner covariance matrices of Gaussian states~\cite{QCB,QCB1,QCB2}, we derive the QCBs for both parties with the details given in Appendix~\ref{App:QCB}.

The resulting QCBs $T_{e,\text{GCS}}^C$ for the GCS at Charlie, $T_{e,\text{GCS}}^E$ for the GCS at Eve,
$T_{e,\text{TMSV}}^C $ for the TMSV at Charlie, and  $T_{e,\text{TMSV}}^E $ for the TMSV at Eve, are
\begin{eqnarray}\label{eq:4QCB}
&&T_{e,\text{GCS}}^C=\frac{1}{2} \left(\inf_{0 \le s \le 1} Q_{s,\text{GCS}}^C \right)^M,\nonumber\\
&&T^{E}_{e,\text{GCS}}=\frac{1}{2} \left(\inf_{0 \le s \le 1} Q_{s,\text{GCS}}^E  \right)^M,\nonumber\\
&&T_{e,\text{TMSV}}^C =\frac{1}{2} \left(\inf_{0 \le s \le 1} Q_{s,\text{TMSV}}^C \right)^M,\nonumber\\
&&T^{E}_{e,\text{TMSV}}=T^{E}_{e,\text{GCS}},
\end{eqnarray}
with the single-copy Chernoff quantities
\begin{eqnarray}
Q_{s,\text{GCS}}^C &=&\frac{1}{(N_B+1)^s (N_1+1)^{1-s}-N_B^s N_1^{1-s}}\;,\nonumber\\
Q_{s,\text{GCS}}^E &=&\frac{1}{(N_B+1)^s (N_E+1)^{1-s}-N_B^s N_E^{1-s}}\;,
\end{eqnarray}
for $N_1=\eta N_s+(1-\eta) N_B$, $N_E=(1-\eta)N_s+\eta N_B$, and $Q_{s,\text{TMSV}}^C$ is omitted for brevity (see Appendix~\ref{App:QCB}). 
In Eq.~(\ref{eq:4QCB}), the equality $T_{e,\text{TMSV}}^E=T_{e,\text{GCS}}^E$ implies that, for the same probe energy $N_s$, the TMSV and GCS probes have the identical covertness against Eve's detection. 
Especially if $N_B=N_s$ for $\eta \neq 1$, one has $N_E=N_B$, $T^E_{e,\text{TMSV}}=T^E_{e,\text{GCS}}=1/2$ achieves the perfect covertness ($\epsilon=0$).
To further quantify the entanglement advantage at the perfect-covertness point, we evaluate the QCB gain
$G_\text{QCB}=10 \log_{10} \left({T_{e,\text{GCS}}^C}/{T_{e,\text{TMSV}}^C} \right)\simeq 0.53(5)dB$ for $\eta=0.01$, $N_B=0.2$, $M=100$, and $\epsilon=10^{-3}$.

\begin{figure}[htp]
	\centering
	\includegraphics[width=0.49\textwidth]{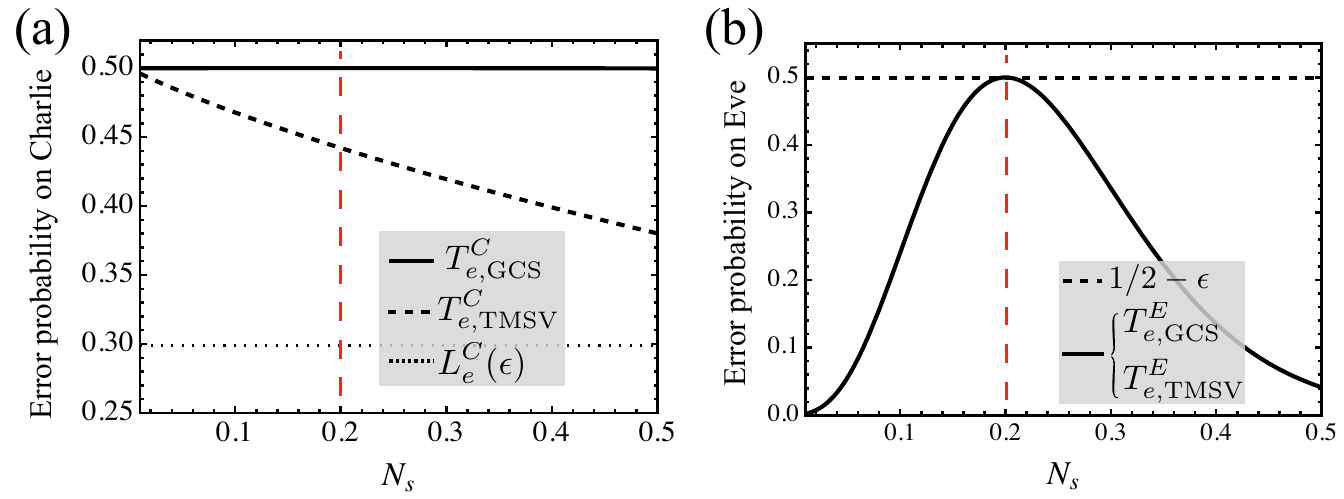}
	\caption
	{
		(a) Error probability on Charlie, including the quantum Chernoff bounds $T_{e,\text{GCS}}^C$ and $T_{e,\text{TMSV}}^C$ for the GCS and TMSV probes, respectively, together with the covertness-induced lower bound $L_e^C$.
		(b) Error probability on Eve, where $T_{e,\text{GCS}}^E=T_{e,\text{TMSV}}^E$, together with the covertness threshold $1/2-\epsilon$.
		The perfect-covertness point $N_s=N_B$ is always marked by the red dashed line.
		The parameters are $\eta=0.01$, $N_B=0.2$, $M=100$, and $\epsilon=10^{-3}$.
	}
	\label{Fig.QCBs}
\end{figure}

We next plot four QCBs of Eq.~(\ref{eq:4QCB}) versus the emissive single-mode mean photon number $N_s$ to illustrate Charlie's discrimination performance and Eve's detectability in Fig.~\ref{Fig.QCBs}.
The covertness-induced lower bound $L_e^{C}$ in Eq.~(\ref{eq:Lc}) on Charlie's discrimination error, and the covertness threshold $1/2-\epsilon$ in Eq.~(\ref{eq:Eveerror}) for Eve's detection error are also exhibited additionally.
As shown in Fig.~\ref{Fig.QCBs} (a), the QCB obtained with the TMSV probe is lower than that induced by the GCS probe, which demonstrates the advantage of signal-idler entanglement for Charlie's target discrimination. 
With the increase of probe energy per mode $N_s$, the error probability caused by the TMSV is further suppressed. 
But the covertness requires that the probe energy cannot be increased arbitrarily so as to prevent the monitor from Eve.
Fig.~\ref{Fig.QCBs} (b) shows that the Eve-side QCBs for the GCS and TMSV probes are identical. 
Both of them attain their maximum value of $1/2$ at the perfect-covertness point.
As $N_s$ deviates from $N_B$, Eve's QCB decreases, and the detectability of Alice's probing is increased.
To summarize, the TMSV probe provides an entanglement-enhanced discrimination advantage for Charlie without sacrificing covertness relative to the GCS probe at the same signal brightness. 

\section{Perturbative covertness-induced energy bound}\label{Sec:bound}
To reveal how covertness constrains the available sensing resources, we translate the indistinguishability requirement at Eve into bounds on the admissible probe energy at Alice.
We denote by $q_n$ the probability mass function (PMF) of total photon number observed by Eve when Alice probes the target, satisfying $\sum_{n=0}^{\infty}q_n=1$, with the expectation $\mathcal{N}_E=\sum_n n q_n$.  
Likewise, $p_n$ represents the PMF of total photon number of M thermal-background modes, with $\sum_{n=0}^\infty p_n=1$ and  $\mathcal{N}_B=\sum_n n p_n=M N_B$. \\

\textbf{Proposition.}
Under the $\epsilon$-covertness condition and in the perturbative
strong-covertness regime, where
$q_n=p_n+\delta q_n$ with $|\delta q_n|/p_n\ll1$, the deviation $\Delta:=N_s-N_B$ of the probe brightness from the thermal-background level satisfies
\begin{eqnarray}\label{eq:interval}
	|\Delta|\lesssim \frac{4}{1-\eta}  \sqrt{\frac{\epsilon  N_B (N_B+1)}{M}}.
\end{eqnarray}
Equivalently, the probe energy lies within
\begin{equation}
	N_s\in[N_{s,\min}(\epsilon),N_{s,\max}(\epsilon)],
\end{equation}
where
\begin{eqnarray}\label{eq:Ns}
N_{s,\max/\min} (\epsilon)=N_B\pm \frac{4}{1-\eta}  \sqrt{\frac{\epsilon  N_B (N_B+1)}{M}}.
\end{eqnarray}
Thus, the $\epsilon$-covertness confines the probe energy to a window centered around the thermal-background mean photon number per mode $N_B$, with a width scaling as $\mathcal{O}(\sqrt{\epsilon/M})$.

\textit{Proof.}
A necessary condition for the $\epsilon$-covertness can be expressed in terms of the Bhattacharyya coefficient~\cite{FDG1999} between these two photon-number distributions,
\begin{eqnarray}\label{eq:fun}
	B\left(\{q_n\},\{p_n\}\right)=\sum_{n=0}^\infty \sqrt{q_n p_n} \ge 1- 2\epsilon,
\end{eqnarray}
with
\begin{eqnarray}
	p_n=\left( \begin{matrix}
		n+M-1\\
		M-1
	\end{matrix}   \right) \frac{N_B^n}{(N_B+1)^{n+M}},
\end{eqnarray}
Eq.~(\ref{eq:fun}) can be derived using the Fuchs-van de Graaf and Cauchy-Schwarz inequalities~\cite{Gu2024}. 
As shown below, this condition can be translated into an admissible range of the probe energy $N_s$.

In the strong-covertness regime $\epsilon\ll1$, Eq.~(\ref{eq:fun})
requires the two photon-number distributions to be close.
We therefore write
\begin{equation}\label{eq:pertu}
	q_n=p_n+\delta q_n,
\end{equation}
where $\delta q_n$ is a small perturbation satisfying
$\sum_{n=0}^{\infty}\delta q_n=0$, $p_n+\delta q_n\ge0$.
For $|\delta q_n|/p_n\ll1$, the Bhattacharyya coefficient can be expanded to second order in $\delta q_n$,
\begin{eqnarray}\label{eq:Bbound}
{\hspace{-0.5cm}}B(\{q_n\},\{p_n\}) \!=\! \sum_{n=0}^\infty p_n \sqrt{1+\frac{\delta q_n}{p_n}}
\simeq 1 \!-\! \frac{1}{8}\sum_{n=0}^\infty \frac{(\delta q_n)^2}{p_n}.
\end{eqnarray} 
Substituting Eq.~(\ref{eq:Bbound}) into Eq.~(\ref{eq:fun}), we obtain
\begin{eqnarray}\label{eq:epsilon}
\epsilon \gtrsim \frac{1}{16} \sum_{n=0}^{\infty} \frac{(\delta q_n)^2}{p_n}.
\end{eqnarray}

We next map the covertness constraint onto the allowed probe energy.
Based on Eq.~(\ref{eq:pertu}), the difference between the mean photon numbers observed by Eve under two hypotheses can be written as
\begin{eqnarray}\label{eq:Delta}
	|\mathcal{N}_E-\mathcal{N}_B|
	&=&
\left|\sum_n n\delta q_n\right|\nonumber\\
	&=&
\left|\sum_n (n-\mathcal{N}_B)\delta q_n\right|\nonumber\\
	&=&
\left|\sum_n\sqrt{p_n}(n-\mathcal{N}_B)\frac{\delta q_n}{\sqrt{p_n}}\right|\nonumber\\
	&\le&
\sqrt{\sum_n p_n(n-\mathcal{N}_B)^2}
\sqrt{\sum_n\frac{(\delta q_n)^2}{p_n}}\nonumber\\
&\lesssim&4\sqrt{\epsilon M N_B(N_B+1)},
\end{eqnarray}
where $\sum_n p_n (n-\mathcal{N}_B)^2=MN_B(N_B+1)$ denotes the variance of the total photon number of the $M$ thermal modes.
The first inequality follows from the Cauchy-Schwarz inequality, while the last step follows from Eq.~(\ref{eq:epsilon}).

The corresponding per-mode energy constraint is therefore
\begin{eqnarray}\label{eq:sss}
\left|
\frac{\mathcal{N}_E}{M}-N_B
\right|
\lesssim
4\sqrt{
	\frac{\epsilon N_B(N_B+1)}{M}
}.
\end{eqnarray}
Next, we map the Eve-side collected single-mode energy into the Alice-side single-mode probe energy according to Eq.~(\ref{eq:rela}), which gives  $\mathcal{N}_E/{M}=(1-\eta)N_s+\eta N_B$. 
Inserting this result into Eq.~(\ref{eq:sss}), we obtain the allowed probe-energy window as shown in Eq.~(\ref{eq:interval}).
We finally propose the upper and lower bounds on the mean probe photon number per mode as shown in Eq.~(\ref{eq:Ns}).
This is the first main result in this work.
\hfill $\blacksquare$

\begin{figure}[!htp]
	\centering
	\includegraphics[width=0.35\textwidth]{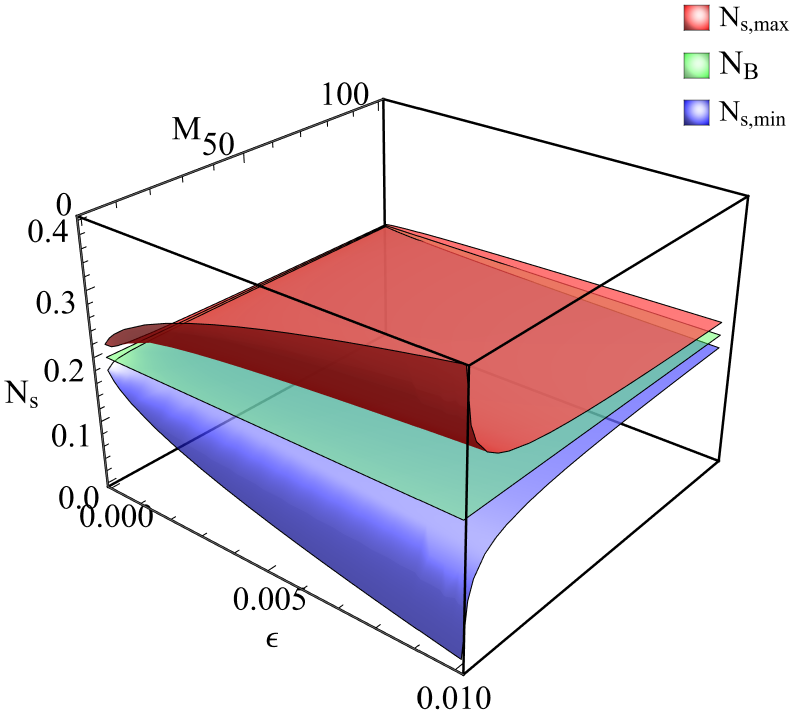}
	\caption
	{Upper bound $N_{s,\max}(\epsilon)$ and lower bound $N_{s,\min}(\epsilon)$ on the mean probe photon number per mode, as functions of the covertness parameter $\epsilon$ and the number of modes $M$.
		The perfect-covertness value $N_s=N_B$ is shown for comparison.
		The parameters are $\eta=0.01$ and $N_B=0.2$.}
	\label{Fig.Nsrange}
\end{figure}

The covertness-induced energy bounds in Eq.~(\ref{eq:Ns}) show that relaxing the covertness requirement (increasing $\epsilon$) or reducing the number of modes broadens the admissible probe-energy window.
This phenomenon is displayed in Fig.~\ref{Fig.Nsrange} by investigating the dependence of $N_{s,\max(\min)}(\epsilon)$ upon $\epsilon$ and $M$.
Notably, this bound is independent of the specific probe-state structure and follows solely from Eve's photon-number statistics, thereby providing a common resource constraint for comparing different quantum probes.
Furthermore, to further assess the accuracy of Eq.~(\ref{eq:Ns}), we determine the admissible range of $N_s$ from Eq.~(\ref{eq:fun}) by using two numerical approaches,  including a Karush-Kuhn-Tucker (KKT)-based optimization~\cite{KKT} and a binary-search (BS) method~\cite{BS}. 
The KKT method mainly optimizes over admissible photon-number PMFs, whereas the BS method focuses on the root finding for the thermal-state family.
In the strong-covertness regime considered here, the numerical results are in good agreement with the analytical prediction, as shown  in Appendix~\ref{App:nume}.

\section{Estimation precision of target reflectivity under the covertness constraint}\label{Sec:CCQFI}
Having established the covertness-constrained bounds for binary target discrimination and probe energy, we now turn to the parameter estimation and investigate how the same covertness requirement limits the metrological precision for the target reflectivity.

\subsection{Quantum Fisher information}
The estimation precision is characterized by the standard deviation $\delta\eta$ of an unbiased estimator of the reflectivity $\eta$.
According to the quantum Cram\'er-Rao bound, one has
\begin{eqnarray}
\delta \eta \ge \frac{1}{\sqrt{\nu F_\eta}},
\end{eqnarray}
where $\nu$ denotes the number of independent experimental repetitions and $F_\eta$ is the quantum Fisher information (QFI) with respect to $\eta$. In the following, we set $\nu=1$.

For Gaussian states, the QFI can be evaluated using several equivalent approaches, including the phase-space formulation based on the first and second moments~\cite{ge,QP,QI7}, the fidelity approach~\cite{E2022,QCB2,Fidelity}, and the spectral decomposition of the density operator~\cite{DS,DS1}. Here, we employ the phase-space method.
For the zero-mean Gaussian states considered in this work, the QFI is given by (see Appendix~\ref{App:QFI})
\begin{eqnarray}\label{eq:QFI}
{\hspace{-0.7cm}}F_\eta=\frac{1}{2} \left( \operatorname{vec}[\partial_\eta\bm V]\right)^T [\bm{V}\otimes\bm{V}-\frac{1}{4}{\bm{\Omega} \otimes \bm{\Omega}}]^{-1} \text{vec}[\partial_\eta \bm{V}],
\end{eqnarray}
where $\bm V$ is the Wigner covariance matrix and $\bm\Omega$ denotes the symplectic form.
Here, $\operatorname{vec}[\bullet]$ denotes the column-wise vectorization of a matrix, obtained by stacking its columns into a single column vector.
We next apply Eq.~(\ref{eq:QFI}) to the GCS and TMSV probes, the relevant derivations are presented in Appendix~\ref{App:QFI}.

Specifically, for Charlie with the GCS probe of Eq.~(\ref{eq:GCS}), the QFI is
\begin{eqnarray}\label{eq:FQ}
	F_\eta^\text{GCS}=\frac{M (N_s-N_B)^2}{N_1 (N_1+1)},
\end{eqnarray}
for $N_1=\eta N_s+(1-\eta)N_B$. 
Here, we have used the additivity of the QFI for product states~\cite{Liu2020}, i.e., $F_Q[\rho^{\otimes M}]=M F_Q[\rho]$.
Especially, at the perfect-covertness point $N_s=N_B \equiv N$, Eq.~(\ref{eq:FQ}) yields $F_\eta^{\mathrm{GCS}}(N)=0$.
This implies that the ensemble-averaged GCS output carries no information about the target reflectivity.

\begin{figure}[htp]
	\centering
	\includegraphics[width=0.3\textwidth]{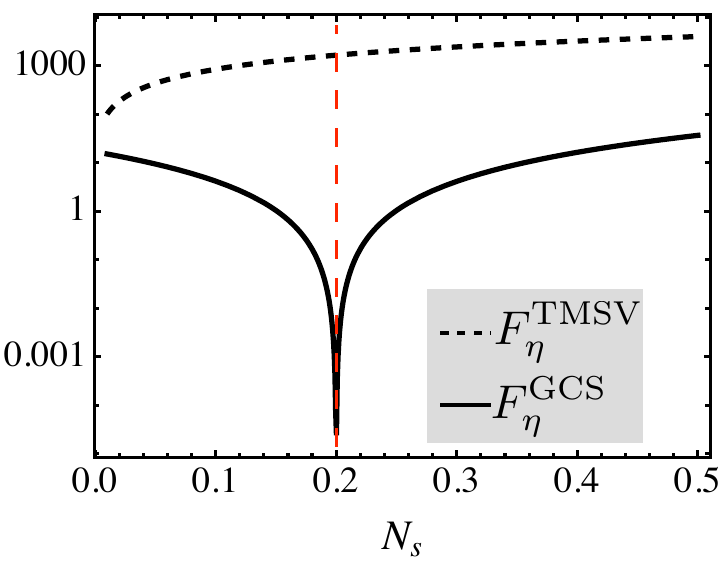}
	\caption
	{ 
	Quantum Fisher informations (QFIs) $F_\eta^{\mathrm{GCS}}$ and $F_\eta^{\mathrm{TMSV}}$ for the GCS and TMSV probes, respectively, plotted on a logarithmic scale.
	The perfect-covertness point $N_s=N_B$ is marked by the red dashed line.
	The parameters are $\eta=0.01$, $N_B=0.2$, and $M=100$.
	}
	\label{Fig.QFIs}
\end{figure}

By contrast, for the TMSV probe in Eq.~(\ref{eq:TMSV}), the QFI is 
\begin{eqnarray}\label{eq:t}
&&F_\eta^{\mathrm{TMSV}}= \frac{M}{\eta(1-\eta)}\nonumber\\
&&{\hspace{1cm}} \times \frac{N_s(N_s+1)+\eta[N_B(2N_s+1)-N_s^2]} {1+(1-\eta)[N_s+N_B+2N_sN_B]}.
\end{eqnarray}
At the perfect-covertness point $N_s=N_B\equiv N$, Eq.~(\ref{eq:t}) reduces to a nonzero value
$F_{\eta}^{\mathrm{TMSV}}(N)\!=\! \frac{M N(N+1)(1+\eta)} {\eta(1-\eta) [1+2N(1-\eta)(1+N)]}$.
Fig.~\ref{Fig.QFIs} compares the QFIs in Eqs.~(\ref{eq:FQ}) and (\ref{eq:t}) as functions of the probe energy $N_s$.
Over the parameter range considered, the TMSV probe exhibits a larger QFI than the GCS probe.
This advantage becomes particularly pronounced at the perfect-covertness point $N_s=N_B$, where the GCS QFI vanishes
whereas the TMSV QFI remains finite. 
This finite TMSV QFI originates from the signal-idler correlations retained by Charlie. 

\subsection{Covertness-constrained quantum Fisher information}
The covertness-induced constraint on the probe energy is given by Eq.~(\ref{eq:interval}).
Under this resource constraint, we define the covertness-constrained quantum Fisher information (CCQFI) as
\begin{eqnarray}
	\mathcal{F}_\eta^{\ell}(\epsilon)= \max_{N_s\in[N_{s,\min}(\epsilon),N_{s,\max}(\epsilon)]} F_\eta^{\ell}(N_s),
\end{eqnarray}
which represents the maximal QFI attainable within the covertness-compatible probe-energy range for $\ell\in\{\mathrm{GCS},\mathrm{TMSV}\}$.

For the GCS state, substituting Eqs.~(\ref{eq:interval}) into (\ref{eq:FQ}) gives
\begin{eqnarray}\label{eq:new}
	F_{\eta}^{\mathrm{GCS}}
	\lesssim
	\frac{
		16\epsilon N_B(N_B+1)
	}{
		N_1(N_1+1)(1-\eta)^2
	},
\end{eqnarray}
with $N_1=\eta N_s+(1-\eta)N_B$.
In the strong-covertness regime, Eq.~(\ref{eq:interval}) implies $N_s=N_B+O(\sqrt{\epsilon/M})$, and hence $N_1=N_B+O(\sqrt{\epsilon/M})\simeq N_B$.
Eq.~(\ref{eq:new}) further yields the GCS CCQFI 
\begin{eqnarray}\label{eq:ep}
	\mathcal F_{\eta}^{\mathrm{GCS}}(\epsilon)
	\simeq
	\frac{16\epsilon}{(1-\eta)^2}   \sim \mathcal{O}(\epsilon).
\end{eqnarray}
It vanishes in the perfect-covertness limit $\epsilon \to 0$ and does not depend on the mode number $M$.

By contrast, the CCQFI of the TMSV probe exhibits a distinct behavior from that of the GCS probe.
Specifically, the QFI in Eq.~(\ref{eq:t}) is monotonically increasing with the probe brightness $N_s$ since ${\partial F_\eta^{\mathrm{TMSV}}}/{\partial N_s}>0$. Hence, the maximum over the covertness-compatible interval is attained at $N_s=N_{s,\max}(\epsilon)$.
Inserting $N_{s,\max}(\epsilon)$ of Eq.~(\ref{eq:Ns}) into Eq.~(\ref{eq:t}) gives the TMSV CCQFI.
In the strong-covertness regime, $\sqrt{\epsilon/M}\ll1$, we expand this CCQFI up to second order in $\sqrt{\epsilon/M}$,
\begin{widetext}
\begin{eqnarray}\label{eq:CCQFITMSVexp}
\mathcal{F}_\eta^{\mathrm{TMSV}}(\epsilon)
&\simeq& \frac{M N_B(N_B+1)(1+\eta)}{\eta(1-\eta)N_\Delta}\nonumber\\
&+&\sqrt{\epsilon M N_B(N_B+1)}\frac{4(2N_B+1)[1+(1-\eta)^2N_B(N_B+1)]}{\eta(1-\eta)^2N_\Delta^2}
-\frac{16\epsilon N_B^2(N_B+1)^2(1+\eta)^2}{\eta(1-\eta)^2N_\Delta^3},
\end{eqnarray}
\end{widetext}
for $N_\Delta=1+2(1-\eta)N_B(N_B+1)$.
Consequently, the asymptotic scaling of the TMSV CCQFI takes the form
\begin{eqnarray}
	\mathcal{F}_\eta^{\mathrm{TMSV}}(\epsilon) \sim  \mathcal{O}(M)+ \mathcal{O}(\sqrt{\epsilon M}) + \mathcal{O}(\epsilon).
\end{eqnarray}

\begin{figure}[htp]
	\centering
	\includegraphics[width=0.3\textwidth]{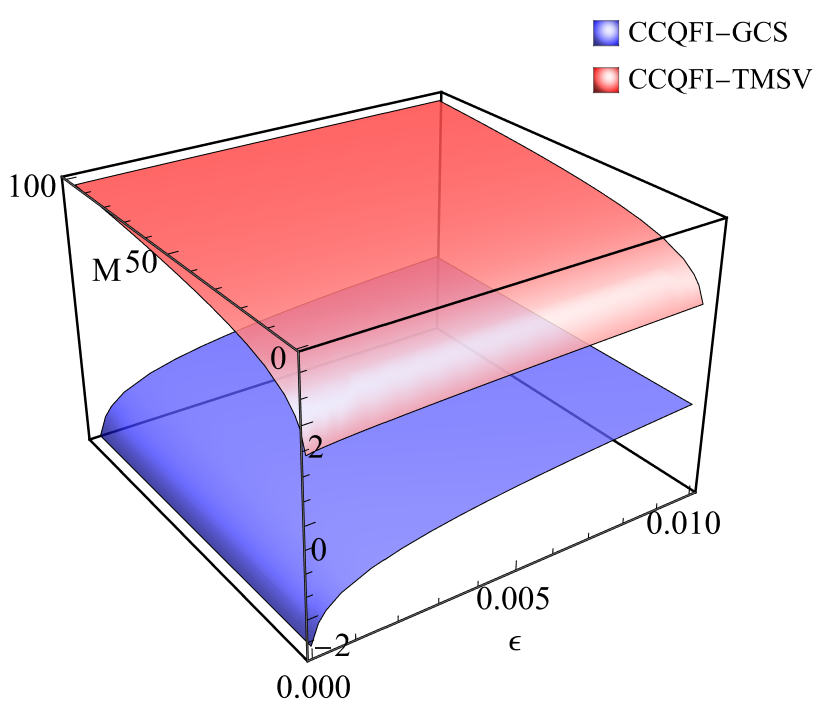}
	\caption
	{ Covertness-constrained quantum Fisher informations (CCQFIs) $\mathcal{F}_\eta^{\mathrm{GCS}}(\epsilon)$ and $\mathcal{F}_\eta^{\mathrm{TMSV}}(\epsilon)$ for the GCS and TMSV probes, respectively, as functions of the covertness parameter $\epsilon$ and the mode number $M$, plotted on a logarithmic scale. 
		The parameters are $\eta=0.01$ and $N_B=0.2$.}
	\label{Fig.CCQFIs}
\end{figure}

These analytical CCQFI scalings constitute the second main result of this work.
Fig.~\ref{Fig.CCQFIs} shows the CCQFIs of the GCS and TMSV probes as functions of the covertness parameter $\epsilon$ and the mode number $M$.
Over the parameter range considered, the TMSV probe consistently yields a larger CCQFI than the GCS probe.
For instance, at $\epsilon=0.001$ and $M=100$, the corresponding CCQFI gain is
$G_{\mathrm{CCQFI}}=
10\log_{10}
\left(
{\mathcal{F}_{\eta}^{\mathrm{TMSV}}}/{\mathcal{F}_{\eta}^{\mathrm{GCS}}}
\right)
\simeq 50.20(1)~\mathrm{dB}$.
Moreover, the TMSV probe can exploit a larger mode number $M$ to enhance the metrological precision while remaining within the covertness-compatible probe-energy window.
In contrast, the GCS CCQFI is independent of $M$ to leading order.

\section{Measurement attainability of the CCQFI}\label{Sec:Meas}
We now investigate the attainability of the CCQFI using two experimentally relevant measurement schemes, homodyne detection (HD) and  photon counting (PC).
For a POVM $\{\Pi_\mathbf{x}| \Pi_\mathbf{x} \ge 0, \sum_\mathbf{x} \Pi_\mathbf{x}=\mathbb{I} \text{(identity operator)}\}$ performed on the parameter-encoded state $\rho_\eta$, the probability distribution of the measurement outcomes $\mathbf{x}$ is $p(\mathbf{x}|\eta)=\mathrm{Tr}[\rho_\eta \Pi_{\mathbf{x}}]$.
The corresponding classical Fisher information (CFI) is
\begin{equation}\label{eq:CFI}
J_\eta=\int d\mathbf{x}\, \frac{[\partial_\eta p(\mathbf{x}|\eta)]^2}{p(\mathbf{x}|\eta)},
\end{equation}
where $\partial_\eta p:=\partial p/\partial \eta$ and the integral is replaced by a summation for discrete measurement outcomes.

The covertness-compatible probe-energy range also constrains the maximal CFI accessible to a given measurement scheme.
We analogously define the covertness-constrained CFI as
\begin{equation}
	\mathcal{J}_{\eta}^{\ell,\mu}(\epsilon)
	=
	\max_{N_s\in[N_{s,\min}(\epsilon),N_{s,\max}(\epsilon)]}
	J_{\eta}^{\ell,\mu}(N_s),
\end{equation}
where $\ell\in\{\mathrm{GCS},\mathrm{TMSV}\}$ labels the probe state and $\mu\in\{\text{HD},\text{PC}\}$ specifies the measurement scheme.
To quantify the fraction of the covertness-constrained metrological information accessible to the specific measurement scheme, we further introduce the attainability ratio
\begin{equation}\label{eq:attainability}
\mathcal{R}^{\ell,\mu}=\frac{\mathcal{J}_{\eta}^{\ell,\mu}(\epsilon)}{\mathcal{F}_{\eta}^{\ell}(\epsilon)},
\qquad
0\leq \mathcal{R}^{\ell,\mu} \leq1.
\end{equation}
The relevant derivations of the CFIs and attainabilities for the homodyne and photon-counting measurements are derived in Appendix~\ref{App:CFI}.

\subsection{Homodyne measurement}
Homodyne detection converts the field quadrature of the returned mode into a measurable photocurrent through interference with a strong phase-reference local oscillator (LO)~\cite{HD,CFI1,CFI2,Kay1993}. 
For Charlie with the GCS probe, the M-copy CFI for the HD scheme is 
\begin{eqnarray}\label{eq:CFI_GCS}
	J_{\eta}^\text{GCS,HD}(N_s)=\frac{M (N_s-N_B)^2}{2\left(\eta N_s+(1-\eta)N_B +\frac{1}{2}\right)^2}.
\end{eqnarray}
Maximizing over the probe-energy-constrained window, the covertness-constrained CFI in the strong-covertness regime is
\begin{equation}\label{eq:GC}
	\mathcal{J}_{\eta}^{\rm GCS, HD}(\epsilon)\simeq
	\frac{8\epsilon N_B(N_B+1)}{(1-\eta)^2\left(N_B+\frac12\right)^2} \sim \mathcal{O}(\epsilon).
\end{equation}
At perfect covertness, this HD measurement cannot distil any information since $\mathcal{J}_{\eta}^{\rm GCS, HD}(0)=0$.
Comparing Eq.~(\ref{eq:GC}) with the GCS CCQFI in Eq.~(\ref{eq:ep}), the homodyne attainability in the strong-covertness limit is
\begin{eqnarray}\label{eq:R1}
	\mathcal{R}^{\rm GCS,HD}=
	\frac{\mathcal{J}_{\eta}^{\rm GCS,HD}(\epsilon)}{\mathcal{F}_{\eta}^{\rm GCS}(\epsilon)}
	=\frac{2N_B(N_B+1)}{(2N_B+1)^2}<1,
\end{eqnarray}
for $\epsilon \neq 0$ and $N_B>0$.

For the TMSV probe, the corresponding covertness-constrained CFI is derived in Appendix~\ref{App:CFI}. Its strong-covertness scaling takes the form
\begin{eqnarray}
	\mathcal{J}_{\eta}^{\rm TMSV, HD}(\epsilon) \sim \mathcal{O}(M)+\mathcal{O}(\sqrt{\epsilon M})+\mathcal{O}(\epsilon),
\end{eqnarray}
which remains finite in the perfect-covertness limit.
The corresponding homodyne attainability takes the form of $\mathcal{R}^{\rm TMSV,HD}={\mathcal{J}_{\eta}^{\rm TMSV,HD}(\epsilon)}/{\mathcal{F}_{\eta}^{\rm TMSV}(\epsilon)}$ with the TMSV CCQFI of Eq.~(\ref{eq:CCQFITMSVexp}).
Its explicit expression is lengthy and is therefore omitted here.

\subsection{Photon counting measurement}
In photon-counting measurements, photon-number-resolving (PNR) detectors~\cite{PC} are used to obtain discrete photon-number outcomes.
For Charlie with the GCS probe, we obtain the $M$-copy CFI for the PC scenario as
\begin{eqnarray}\label{eq:GCSCFI}
	J_\eta^{\rm GCS,PC}(N_s)=\frac{M(N_s-N_B)^2}{N_1(N_1+1)}.
\end{eqnarray}
This is identical to the corresponding GCS QFI in Eq.~(\ref{eq:FQ}).
Hence, maximizing both quantities over the same covertness-induced probe-energy window gives
\begin{eqnarray}\label{eq:GCSHD}
	&&\mathcal J_\eta^{\rm GCS,PC}(\epsilon) \simeq \frac{16\epsilon}{(1-\eta)^2}   \sim \mathcal{O}(\epsilon), \nonumber\\
	&&\mathcal R^{\rm GCS,PC}=\frac{\mathcal J_\eta^{\rm GCS,PC}(\epsilon)}{\mathcal F_\eta^{\rm GCS}(\epsilon)}=1,
\end{eqnarray}
for $\epsilon>0$.
It follows that photon counting saturates the GCS QFI and constitutes an optimal measurement for estimating the target reflectivity.

For the TMSV state, the corresponding CFI and its covertness-constrained form in the strong-covertness regime are evaluated numerically in Appendix~\ref{App:CFI} using suitably truncated photon-number sums.
Its asymptotic scaling takes the form 
\begin{eqnarray}\label{eq:QFIscaling}
	\mathcal{J}_{\eta}^{\rm TMSV, PC}(\epsilon) \sim \mathcal{O}(M)+\mathcal{O}(\sqrt{\epsilon M})+\mathcal{O}(\epsilon).
\end{eqnarray}
The resulting PC attainability $\mathcal{R}^{\rm TMSV,PC}={\mathcal{J}_{\eta}^{\rm TMSV,PC}(\epsilon)}/{\mathcal{F}_{\eta}^{\rm TMSV}(\epsilon)}$ is analytically cumbersome and is evaluated numerically.

\begin{figure}[!htp]
	\centering
	\includegraphics[width=0.5\textwidth]{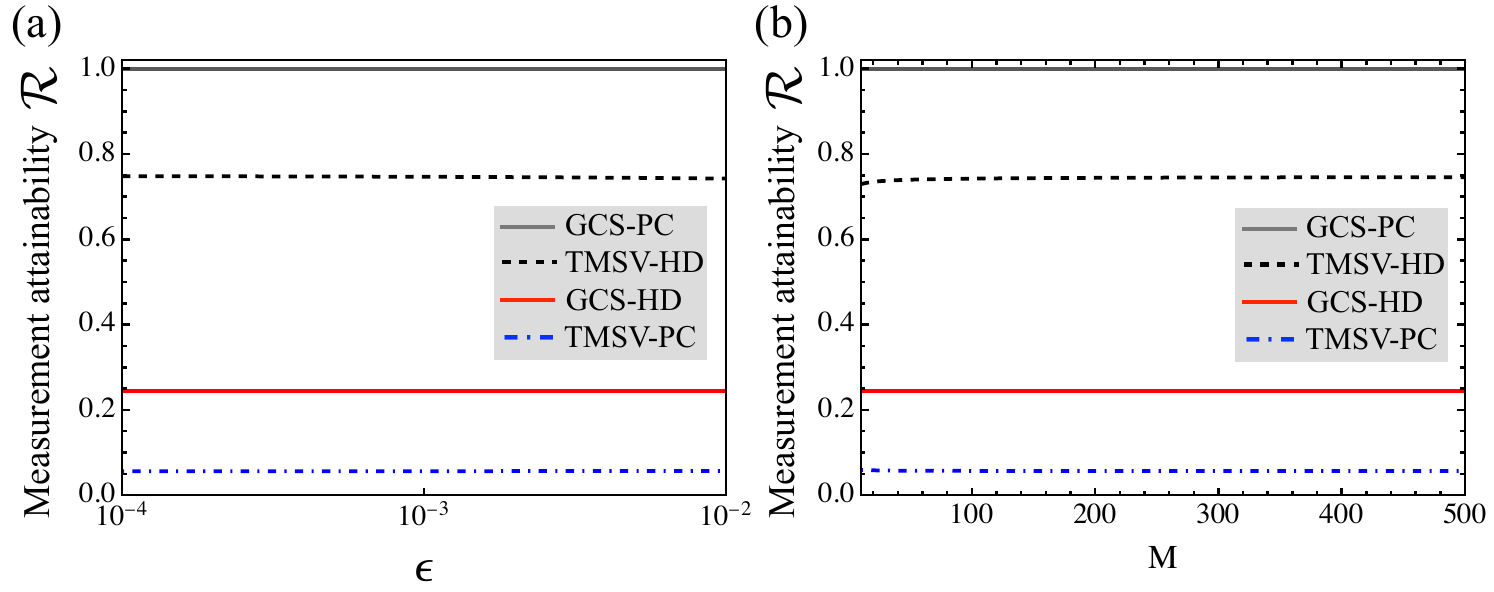}
	\caption
	{Measurement attainabilities
		$\mathcal{R}^{\rm GCS,HD}$,
		$\mathcal{R}^{\rm TMSV,HD}$,
		$\mathcal{R}^{\rm GCS,PC}$, and
		$\mathcal{R}^{\rm TMSV,PC}$:
		(a) as functions of the covertness parameter $\epsilon$ and
		(b) as functions of the mode number $M$.
		The parameters are $\eta=0.01$ and $N_B=0.2$. In panel (a) $M=100$, while in panel (b) $\epsilon=10^{-2}$.}
	\label{Fig.R}
\end{figure}

Finally, we compare the four measurement attainabilities, $\mathcal{R}^{\rm GCS,HD}$, $\mathcal{R}^{\rm TMSV,HD}$, $\mathcal{R}^{\rm GCS,PC}$, and $\mathcal{R}^{\rm TMSV,PC}$ in Fig.~\ref{Fig.R}. For the parameters considered here, their attainabilities follow the ordering
\begin{equation}
	\mathcal{R}^{\rm GCS,PC}
	>
	\mathcal{R}^{\rm TMSV,HD}
	>
	\mathcal{R}^{\rm GCS,HD}
	>
	\mathcal{R}^{\rm TMSV,PC}.
\end{equation}
In particular, the photon counting saturates the GCS CCQFI, whereas joint homodyne detection provides a higher attainability than joint photon counting for the TMSV probe in the parameter regime considered.

\section{Conclusion}\label{Sec:conclusion}
In this work, we have established a unified framework connecting covert target discrimination, covertness-constrained sensing resources, and quantum-enhanced parameter estimation.
By investigating the quantum Chernoff bounds in target discrimination, we have shown that the TMSV probe outperforms the GCS probe at the same signal brightness while preserving the same covertness.
We have further established perturbative $\epsilon$-covertness-induced bounds on the probe energy, which confine the admissible brightness to a window centered at $N_B$ with width scaling as $\mathcal{O}(\sqrt{\epsilon/M})$.
Moreover, at the perfect-covertness point, the TMSV probe retains a finite QFI, whereas the GCS QFI vanishes, revealing a pronounced metrological advantage enabled by signal-idler correlations.
Additionally, to incorporate the covertness constraint directly into parameter estimation, we have introduced the covertness-constrained quantum Fisher information.
Within the covertness-compatible energy window, the GCS CCQFI scales as $\mathcal{O}(\epsilon)$, whereas the TMSV CCQFI exhibits the asymptotic scaling $\mathcal{O}(M)+\mathcal{O}(\sqrt{\epsilon M})+\mathcal{O}(\epsilon)$. 
In the strong covertness regime $\epsilon \to 0$ at fixed $M$, the GCS CCQFI vanishes, whereas the TMSV CCQFI retains a finite $\mathcal{O}(M)$ contribution, revealing a pronounced metrological advantage.
Finally, we further assess the attainability of these limits through covertness-constrained CFIs for homodyne and photon-counting measurements.
Photon counting saturates the GCS CCQFI, whereas homodyne detection achieves a higher attainability than photon counting for the TMSV probe.

In terms of experimental feasibility, quantum illumination has been experimentally demonstrated across several platforms, including optical and microwave implementations~\cite{QI2,Barzanjeh2020,Xu2021}.
Entanglement-enhanced covert sensing has also been experimentally demonstrated on an optical platform~\cite{E2022}.
These advances provide a promising pathway toward realizing covert target sensing within the covertness-constrained framework developed here.
Another important direction is to extend the present single-target setting to multiparameter sensing~\cite{QI7}. This requires jointly optimizing the estimation of multiple target parameters under a global covertness constraint, thereby introducing a new trade-off between multiparameter sensing precision and adversarial detectability.

\begin{acknowledgments}
Fengyu Guan has the same contributions for this work with Yu Yang. 
We thanks for the useful discussions with Matteo G. A. Paris.
This research is supported by the Fundamental Research Funds for the Central Universities (Grant No. xxj032025044),
National Natural Science Foundation of China (NSFC) (Grants Nos. 12204371 and 12474363),
Shaanxi Fundamental Science Research Project for Mathematics and Physics (Grants No. 23JSQ012).
\end{acknowledgments}

\section*{DATA AVAILABILITY}
The data that support the findings of this article are not publicly available. The data are available from the authors upon reasonable request.

\appendix

\section{Quantum Chernoff bound with mean vector and Wigner covariance matrix}\label{App:QCB}
For an arbitrary $n$-mode bosonic system, we define the quadrature vector as $\mathbf{Q}=(q_1,\cdots q_n,p_1,\cdots p_n)^T$, where $q_\ell=\frac{a_\ell+a_\ell^\dagger}{\sqrt{2}}$ and $p_\ell=\frac{-i(a_\ell-a_\ell^\dagger)}{\sqrt{2}}$ are associated with the bosonic annihilation and creation operators $a_\ell$ and $a_\ell^\dagger$ for $\ell=1,\cdots,n$. 
The components of $\mathbf{Q}$ satisfy the commutation relation 
\begin{eqnarray}
	[Q_j,Q_k]=i\Omega_{jk},
\end{eqnarray}
where the elements $\{\Omega_{jk}\}_{j,k=1,\cdots, n}$ constitute a symplectic form
\begin{eqnarray}
\mathbf{\Omega}=\left(\begin{matrix}
\mathbf{0} & \mathbf{I}_{n \times n}\\
-\mathbf{I}_{n \times n} & \mathbf{0}
\end{matrix}\right).
\end{eqnarray}
Here, $\mathbf{0}$ and $\mathbf{I}$ denote a zero matrix and a dentity matrix with $n \times n$ size, respectively.

A Gaussian state $\rho$ can be characterized by its mean vector (i.e., the first moment of $\bm{Q}$)
\begin{eqnarray}\label{eq:av}
	\bar{\mathbf{Q}}=\operatorname{Tr}(\rho\mathbf{Q}),
\end{eqnarray}
and Wigner covariance matrix (i.e., the second central moment of $\bm{Q}$)
\begin{eqnarray}\label{eq:cov}
	V_{jk}
	=
	\frac{1}{2}
	\operatorname{Tr}
	\left[
	\rho
	\left(
	\Delta Q_j\Delta Q_k+
	\Delta Q_k\Delta Q_j
	\right)
	\right],
\end{eqnarray}
where $\Delta Q_j=Q_j-\bar{Q}_j$.
With this convention, the covariance matrix for the vacuum state is $\bm{V}_{\rm vac}=\mathbf{I}_{2n\times 2n}/2$.

According to the Williamson theorem～\cite{QCB1}, for every $\mathbf{V}$ one can find a $2n$-dimensional symplectic matrix $\mathbf{S}$ satisfying $\mathbf{S} \mathbf{\Omega} \mathbf{S}^T=\mathbf{\Omega}$.  The Wigner covariance matrix can be decomposed as
\begin{eqnarray}
\mathbf{V}=\mathbf{S} 
\begin{pmatrix}
\bm{\nu}& \mathbf{0} \\
	\mathbf{0} & \bm{\nu}
\end{pmatrix}
 \mathbf{S}^T, \quad \bm{\nu}=\text{diag}(\nu_1,\cdots,\nu_n).
\end{eqnarray}
The quantities $\nu_\ell$ are symplectic eigenvalues for $\ell=1,\cdots, n$ and constitute the symplectic spectrum of $\mathbf{V}$. 
The eigenvalues of $i\mathbf{\Omega} \mathbf{V}$ occur in pairs $\{\pm \nu_1,\cdots,\pm\nu_n\}$~\cite{Gaussian_Review}.

The Quantum Chernoff bound associated with single-copy hypothesis states is defined as
\begin{eqnarray}\label{eq:Q}
	T_e=\frac{1}{2}e^{-\xi}=\frac{1}{2} \inf_{0 \le s \le 1} Q_s,\;\; Q_s=\operatorname{Tr}[\rho_0^s \rho_1^{1-s}],
\end{eqnarray}
where $\xi=-\ln \left(\inf_{0 \le s \le 1} Q_s\right)$ denotes the Quantum Chernoff  exponent.
Here $\rho_0$ and $\rho_1$ corresponds to the target-absence and target-presence hypothesis states with single copy, respectively.
The variable $s \in [0,1]$ is the Chernoff parameter, and $s=1/2$ gives the quantum Bhattacharyya bound (QB) $T_e^\text{QB}$ satisfying $T_e \le T_e^\text{QB}$.
Besides, for M independent copies of hypothesis states, the QCB becomes
\begin{eqnarray}
T_e^{(M)}=\frac{1}{2} e^{-M \xi}=\frac{1}{2} \left(\inf_{0 \le s \le 1} Q_s\right)^M.
\end{eqnarray}

Following the Williamson decomposition, we introduce two auxiliary functions
\begin{eqnarray}
	g_s(\nu)&=&\frac{1}{(\nu+\frac{1}{2})^s-(\nu-\frac{1}{2})^s},\nonumber\\
	\lambda_s(\nu)&=&\frac{1}{2} \frac{(\nu+\frac{1}{2})^s+(\nu-\frac{1}{2})^s}{(\nu+\frac{1}{2})^s-(\nu-\frac{1}{2})^s}.
\end{eqnarray}
The symplectic spectra of the two hypothesis states are denoted by $\{\nu_{0,k}\}_{k=1}^{n}$ and $\{\nu_{1,k}\}_{k=1}^{n}$,
respectively. 
The Chernoff quantity $Q_s$ of Eq.~(\ref{eq:Q}) can then be expressed in terms of the mean vectors and Wigner covariance matrices as
\begin{eqnarray}\label{eq:Qs}
Q_s=X \exp(-\frac{1}{2} \mathbf{d}^T [\mathbf{V}_0(s)+\mathbf{V}_1(1-s)]^{-1}  \mathbf{d}),
\end{eqnarray}
where $\mathbf{d} = \operatorname{Tr}[\rho_0 \mathbf{Q} ] - \operatorname{Tr}[\rho_1 \mathbf{Q}] $ and
\begin{eqnarray}\label{eq:rele}
&&X = \frac{ \prod_{k=1}^n   g_s(\nu_{0,k})g_{1-s}(\nu_{1,k}) }{\sqrt{\operatorname{det}[\mathbf{V}_0(s)+\mathbf{V}_1(1-s)   ]}},\nonumber\\
&&\mathbf{V}_0(s)
= \mathbf{S}_0
\begin{pmatrix}
\bm{\lambda}_0 & \mathbf{0} \\ 
	\mathbf{0} & 	\bm{\lambda}_0 
\end{pmatrix}
\mathbf{S}_0^T, \nonumber\\
&&\bm{\lambda}_0=\text{diag}(\lambda_s(\nu_{0,1}), \cdots, \lambda_s(\nu_{0,n})), \nonumber\\
&&\mathbf{V}_1 (1-s) 
= \mathbf{S}_1
\begin{pmatrix}
\bm{\lambda}_1 & \mathbf{0} \\ 
	\mathbf{0} &  \bm{\lambda}_1
\end{pmatrix}
\mathbf{S}_1^T, \nonumber\\
&& \bm{\lambda}_1=\text{diag}\left(\lambda_{1-s}(\nu_{1,1}),\cdots, \lambda_{1-s}(\nu_{1,n})\right).
\end{eqnarray}

We then apply the above formalism to the $M$ i.i.d. copies of GCS and TMSV probes and derive the corresponding QCBs.
\paragraph{Gaussian-distributed coherent-state probe.}
For each copy of the GCS probe  (\ref{eq:GCS}), only one signal mode is involved, i.e., $\mathbf{Q}=(q,p)^T$ with $n=1$.
For Charlie, the single-copy hypothesis states are  
\begin{eqnarray}
{\hspace{-1cm}} \rho_{0,C}^{(1)}&=&\rho_\text{th}^{(m)}(N_B) \nonumber\\
{\hspace{-1cm}} \rho_{1,C}^{(1)}&=& \rho_\text{th}^{(m)}(\eta N_s+(1-\eta)N_B),
\end{eqnarray}
where $\rho_\text{th}^{(m)}(\bullet)$ is defined in Eq.~(\ref{eq:back}).
Based on Eqs.~(\ref{eq:av})-(\ref{eq:cov}), the mean vector and Wigner covariance matrix for different hypothesis states are given by
\begin{eqnarray}\label{eq:QVC}
\bar{\mathbf{Q}}_0&=&\bar{\mathbf{Q}}_1=\left( \begin{matrix} 0\\0 \end{matrix} \right),  \quad
\mathbf{V}_0=\left(N_B+\frac{1}{2}\right) \mathbf{I}_{2 \times 2},\nonumber\\
\mathbf{V}_1 &=& \left( \eta N_s+(1-\eta)N_B+\frac{1}{2 }\right) \mathbf{I}_{2 \times 2}.
\end{eqnarray}
We further get $\mathbf{d}=0$,  $\nu_{0,1}=N_B+\frac{1}{2}$, $\nu_{1,1}=\eta N_s+(1-\eta)N_B+\frac{1}{2}$, $\mathbf{S}_0=\mathbf{S}_1=\mathbf{I}_{2 \times 2}$.
Then we have $\bm{\lambda}_0=\lambda_s(\nu_{0,1})=\frac{1}{2} \frac{(N_B+1)^s+N_B^s}{(N_B+1)^s-N_B^s}$, $\bm{\lambda}_1=\lambda_{1-s}(\nu_{1,1})=\frac{1}{2} \frac{(N_1+1)^{1-s}+N_1^{1-s}}{(N_1+1)^{1-s}-N_1^{1-s}}$ for $N_1=\eta N_s+(1-\eta)N_B$.
One further gets
$\mathbf{V}_0(s)=\lambda_s(\nu_{0,1}) \mathbf{I}_{2\times 2}$, 
$\mathbf{V}_1(1-s)=\lambda_{1-s}(\nu_{1,1}) \mathbf{I}_{2 \times 2}$, 
and $X=\frac{g_s(\nu_{0,1}) g_{1-s}(\nu_{1,1})}{\lambda_s(\nu_{0,1})+\lambda_{1-s}(\nu_{1,1})}$ with $g_s({\nu_{0,1}})=\frac{1}{(N_B+1)^s-N_B^s}$ and $g_{1-s}({\nu_{1,1}})=\frac{1}{(N_1+1)^{1-s}-N_1^{1-s}}$.
Finally, we have 
$Q_{s,\text{GCS}}^C=X=\frac{1}{(N_B+1)^s (N_1+1)^{1-s}-N_B^s N_1^{1-s}}$.
The minimization of $Q_{s,\text{GCS}}^C$ over $s\in[0,1]$ is performed numerically. 
The resulting M-copy QCB is $T^{C}_{e,\text{GCS}}=\frac{1}{2} \left(\inf_{0 \le s \le 1} Q_{s,\text{GCS}}^C \right)^M$.

For Eve, the same procedure gives the single-copy hypothesis states as
\begin{eqnarray}\label{eq:s1}
\rho_{0,E}^{(1)}&=&\rho_\text{th}^{(m)} (N_B)\;,\nonumber\\
\rho_{1,E}^{(1)}&=&\rho_\text{th}^{(m)}\left[(1-\eta)N_s+\eta N_B \right].
\end{eqnarray}
The corresponding mean vectors and Wigner covariance matrices are
\begin{eqnarray}
\bar{\mathbf{Q}}_{0}&=&\bar{\mathbf{Q}}_{1}=\left(\begin{matrix} 0\\ 0 \end{matrix}\right), \mathbf{V}_0=\left(N_B+\frac{1}{2}\right) \mathbf{I}_{2 \times 2},\nonumber\\
\mathbf{V}_1&=&\left((1-\eta)N_s+\eta N_B+\frac{1}{2} \right) \mathbf{I}_{2 \times 2}.
\end{eqnarray}
The relevant qiantities in Eq.~(\ref{eq:rele}) are obtained correspondingly, which yields $Q_{s,\text{GCS}}^E=\frac{1}{(N_B+1)^s (N_E+1)^{1-s}-N_B^s N_E^{1-s}}$ for $N_E=(1-\eta)N_s+\eta N_B$.
Minimizing $Q_{s,\text{GCS}}^E$ over $s\in[0,1]$ yields the M-copy QCB $T^{E}_{e,\text{GCS}}=\frac{1}{2} \left(\inf_{0 \le s \le 1} Q_{s,\text{GCS}}^E \right)^M$.

\paragraph{Two-mode squeezed-vacuum probe.}
For each copy of the TMSV probe (see Eq.~(\ref{eq:TMSVsingle})), the idler and return modes are jointly used so that $\mathbf{Q}=(q_I,q_R,p_I,p_R)^T$ with $n=2$. Here the subscripts $I$ and $R$ denote the ilder and return modes.
For Charlie, the single-copy hypothesis states are 
\begin{eqnarray}
\rho_{0,C}^{(1)}&=&\operatorname{Tr}_S	\left[|\psi_{\rm TMSV}\rangle \langle\psi_{\rm TMSV}|\right]
\otimes \rho_\text{th}^{(m)}(N_B) \nonumber\\
&=&\rho_\text{th}^{(m)}(N_s) \otimes  \rho_\text{th}^{(m)}(N_B),  \nonumber\\
 \rho_{1,C}^{(1)}&=& \operatorname{Tr}_E[(\mathbb{I} \otimes U_{BS}^{(m)}) ([|\psi_{\rm TMSV}\rangle \langle\psi_{\rm TMSV}| \nonumber\\
&& \otimes \rho_\text{th}^{(m)}(N_B)) ( \mathbb{I} \otimes U_{BS}^{(m)})^\dagger],
\end{eqnarray}
where $\mathbb{I}$ denotes the identity operator acting on the idler mode.
Similarly, based on Eqs.~(\ref{eq:av})-(\ref{eq:cov}), the mean vector and Wigner covariance matrix for different hypothesis states are written into
\begin{eqnarray}\label{eq:ss}
\bar{\mathbf{Q}}_0&=&\bar{\mathbf{Q}}_1=\left(\begin{matrix}  0\\0\\0\\0    \end{matrix}\right), \mathbf{V}_0=\left(\begin{matrix} B & 0\\ 0 & D \end{matrix}\right) \oplus \left(\begin{matrix} B & 0\\ 0 & D \end{matrix}\right),  \nonumber\\
\mathbf{V}_1&=&\left(\begin{matrix} B & C\\ C & A \end{matrix}\right) \oplus \left(\begin{matrix} B& -C\\ -C & A \end{matrix}\right),
\end{eqnarray}
where $A=\eta N_s+(1-\eta)N_B+1/2$, $B=N_s+1/2$, $C=\sqrt{\eta N_s (N_s+1)}$, and $D=N_B+1/2$.
We further get $\mathbf{d}=0$, $\{\nu_{0,1},\nu_{0,2}\}=\{B,D\}$ and $\{\nu_{1,1},\nu_{1,2}\}=\{\frac{R+B-A}{2}, \frac{R-B+A}{2}\}$ for $R=\sqrt{(A+B)^2-4C^2}$.

Additionally, for the matrix $\mathbf{V}_0(s)$, the relevant matrices are $\mathbf{S}_0=\mathbf{I}_{4\times 4}$ and $\bm{\lambda}_0=\text{diag}(\lambda_s(\nu_{0,1}),\lambda_s(\nu_{0,2}))$.
For the matrix $\mathbf{V}_1(1-s)$, the relevant matrices are $\mathbf{S}_1=\bm{R}_1 \oplus \bm{R}_2$ for 
$\bm{R}_1=\left( \begin{matrix} 
\cosh x & \sinh x \\ \sinh x & \cosh x
\end{matrix}\right)$ 
and $\bm{R}_2=\left( \begin{matrix} 
\cosh x & -\sinh x \\ -\sinh x & \cosh x
\end{matrix}\right)$ with $\tanh(2x)=\frac{2C}{A+B}$, and  $\bm{\lambda}_1=\text{diag}(\lambda_{1-s}(\nu_{1,1}), \lambda_{1-s}(\nu_{1,2}))$.
It follows that $Q_{s,\text{TMSV}}^C=X$ is obtained by substituting these results into (\ref{eq:Qs}),
and the resulting M-copy QCB is $T^{C}_{e,\text{TMSV}}=\frac{1}{2} \left(\inf_{0 \le s \le 1} Q_{s,\text{TMSV}}^C \right)^M$.

For Eve, the single-copy hypothesis states are
\begin{eqnarray}\label{eq:s2}
\rho_{0,E}^{(1)}&=&\rho_{\rm th}^{(m)}(N_B),\nonumber\\
\rho_{1,E}^{(1)}&=&
\operatorname{Tr}_R
	[U_{BS}^{(m)}
	(\operatorname{Tr}_I[
	|\psi_{\rm TMSV}\rangle
	\langle\psi_{\rm TMSV}|] \nonumber\\
	&&\otimes
	\rho_{\rm th}^{(m)}(N_B))
	( U_{BS}^{(m)})^\dagger]
	\nonumber\\
	&=&
	\rho_{\rm th}^{(m)}
	\left[
	(1-\eta)N_s+\eta N_B
	\right].
\end{eqnarray}
Tracing out the idler of the $m$-th TMSV pair gives
\begin{eqnarray}
	\operatorname{Tr}_I
	\left[
	|\psi_{\rm TMSV}\rangle
	\langle\psi_{\rm TMSV}|
	\right]
	=
	\rho_{\rm th}^{(m)}(N_s).
\end{eqnarray}
Comparing Eqs.~(\ref{eq:s1}) and (\ref{eq:s2}), Eve observes identical single-copy hypothesis states for the TMSV and GCS probes with the same $N_s$. Since the $M$-copy states are $\rho_{j,E}=\left(\rho_{j,E}^{(1)}\right)^{\otimes M}$, their QCBs are therefore equal, i.e., $T_{e,\rm GCS}^E=T_{e,\rm TMSV}^E$.

\section{Numerical verification of the restricted probe energy}\label{App:nume}
\begin{figure}[!htp]
	\centering
	\includegraphics[width=0.5\textwidth]{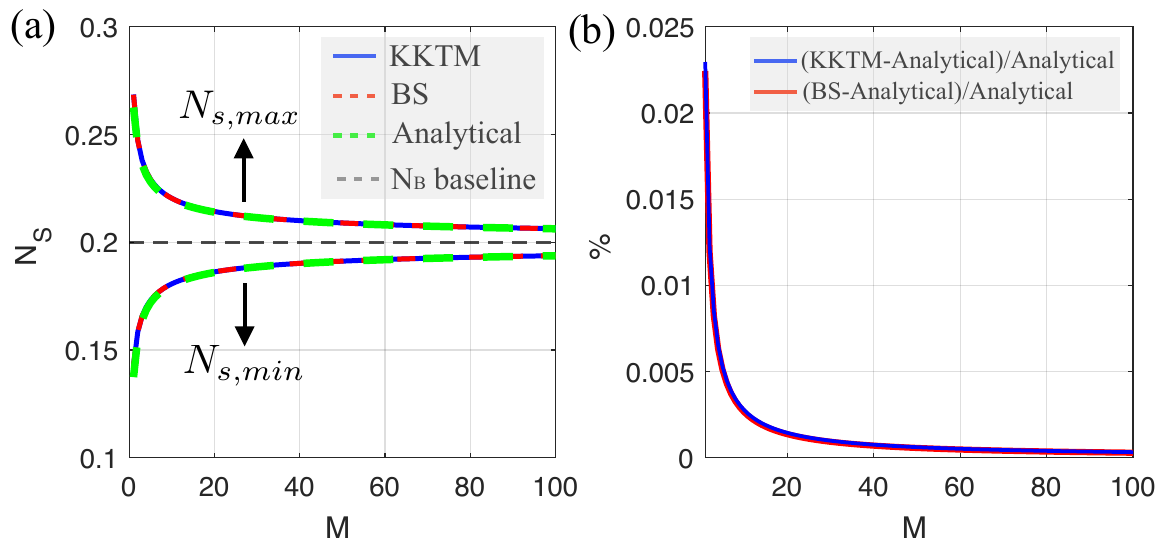}
	\caption
	{(a) Upper bound $N_{s,\max}(\epsilon)$ and lower bound $N_{s,\min}(\epsilon)$ on the mean probe photon number per mode as functions of the copy number of quantum state $M$ under the covertness constraint. The results obtained from the KKT-based method and the BS approach are compared with the analytical prediction in Eq.~(\ref{eq:Ns}).		
		(b) Relative errors of the numerical results with respect to the analytical prediction.
		The parameters $\eta=0.01$, $N_B=0.2$, and $\epsilon=10^{-3}$ are set.}
	\label{Fig.com}
\end{figure}

The covertness-induced constraint on the probe energy per mode is analytically derived in Eq.~(\ref{eq:Ns}), which is a perturbative strong-covertness approximation.
To further assess the accuracy of this analytical result, we also determine the allowed range of $N_s$ from Eq.~(\ref{eq:fun}) by using two numerical approaches, including a Karush-Kuhn-Tucker (KKT)-based optimization and a binary-search (BS) method. 
The pseudocodes for the KKT and BS approaches are presented below.
The analytical and numerical results are compared in Fig.~\ref{Fig.com}, which shows a good agreement. 
\begin{figure}[!htp]
	\centering
	\begin{minipage}{0.95\columnwidth}
		\begin{center}
		\textbf{ALGORITHM 1. KKT optimization}
		\end{center}
		\hrule
		\vspace{2pt}
		\hrule
		\vspace{5pt}
		\begin{algorithmic}[1]
			\Require Background brightness $N_B$, 
			the copy number of probe state $M$,
			covertness constraint $\epsilon$, 
			reflectivity $\eta$,
			and a sufficiently large photon-number cutoff $d$.
			\Ensure $N_{s,\max}$ and $N_{s,\min}$.
		
			\State Extremise the Eve-side energy objective $f (\{q_n\})=\sum_{n=0}^{d} n\,q_n$.
			
			\State Define the normalization constraint $h(\{q_n\})=\sum_{n=0}^{d} q_n - 1 = 0$ and the covertness inequality $g(\{q_n\})=1-2\epsilon-\sum_{n=0}^d \sqrt{q_n p_n}\le 0$. Here, $p_n$ is the total-photon-number PMF of the
			$M$ thermal-background modes, which is renormalized after truncation to $d$.

	        \For{$\ell\in\{-,+\}$}

			\State Set $f_-=f$ for the lower-energy branch and $f_+=-f$ for the upper-energy branch.

			\State Build the Lagrangian function
			$\mathcal L=
			f_\ell(\{q_n\})
			+\lambda_1 g(\{q_n\})
			+\lambda_2 h(\{q_n\})$.
			
			\State \textbf{Stationarity:} solve $\partial \mathcal{L}/\partial q_n = 0$ to obtain $q_n^{*(\ell)}(\lambda_1,\lambda_2)$.

			\State \textbf{Primal feasibility:} require $q_n^{*(\ell)}\ge0$, $\sum_{n=0}^{d}q_n^{*(\ell)}=1$, and $g(\{q_n^{*(\ell)}\})\le0$.
			
			\State \textbf{Dual feasibility:} require $\lambda_1 \ge 0$.
			
			\State \textbf{Complementary slackness:} $\lambda_1g(\{q_n^{*(\ell)}\})=0$. Since stationarity requires $\lambda_1>0$, complementary slackness implies $g(\{q_n^{*(\ell)}\})=0$.
			
           \State Solve the normalization condition together with the saturated-covertness condition for $\lambda_1$ and $\lambda_2$ using a numerical root finder with an absolute tolerance of $10^{-10}$.
           
			\State Compute the corresponding per-mode probe brightness $N_s^\ell=\frac{f(\{q_n^{*(\ell)}\})/M-\eta N_B}{1-\eta}$.
				        
			\EndFor
			
			\State \Return $N_{s,\max}=N_s^+$ and $N_{s,\min}=N_s^-$.
		\end{algorithmic}
		\vspace{5pt}
		\hrule
		\vspace{2pt}
		\hrule
	\end{minipage}
\end{figure}

\begin{figure}[!htp]
	\centering
	\begin{minipage}{0.95\columnwidth}
		\begin{center}
			\textbf{ALGORITHM 2. BS method}
		\end{center}
		\hrule
		\vspace{2pt}
		\hrule
		\vspace{5pt}
		\begin{algorithmic}[1]
			\Require Background brightness $N_B$,
			copy number of probe state $M$,
			covertness parameter $\epsilon$,
			reflectivity $\eta$,
			a sufficiently large photon-number cutoff $d$,
			and numerical tolerance $\mathrm{Tol}=10^{-10}$.

			\Ensure $N_{s,\max}$ and $N_{s,\min}$.
			
			\State Set the covertness threshold $c=1-2\epsilon$.
			
			\State For a trial probe brightness $N_s$, define the total-photon-number PMFs
$p_n=\binom{n+M-1}{M-1}\frac{N_B^n}{(N_B+1)^{n+M}}$ and
$q_n(N_s)=\binom{n+M-1}{M-1}\frac{N_E(N_s)^n}{[N_E(N_s)+1]^{n+M}}$ with $N_E(N_s)=(1-\eta)N_s+\eta N_B$, and 
renormalize the truncated PMFs over $n=0,\ldots,d$. 
			
			\State Define the covertness score (i.e., Bhattacharyya coefficient) $L(N_s)=\sum_{n=0}^{d}\sqrt{p_nq_n(N_s)}$.
			
			\Statex

			\State \textbf{Upper branch: determination of $N_{s,\max}$.}
			
			\State Set $N_{\rm in}=N_B$, for which $L(N_{\rm in})\ge c$.
			
			\State Increase a trial endpoint $N_{\rm out}>N_B$ until
			$L(N_{\rm out})<c$.
			
			\While{$|N_{\rm out}-N_{\rm in}|>\mathrm{Tol}$}
			
			\State Set
$N_{\rm mid}=\frac{N_{\rm in}+N_{\rm out}}{2}$.
			
			\If{$L(N_{\rm mid})\ge c$}
			\State $N_{\rm in}\gets N_{\rm mid}$.
			\Else
			\State $N_{\rm out}\gets N_{\rm mid}$.
			\EndIf
			
			\EndWhile
			
			\State Set $N_{s,\max}=\frac{N_{\rm in}+N_{\rm out}}{2}$.
			
			\Statex
			
			\State \textbf{Lower branch: determination of $N_{s,\min}$.}
			
			\State Set $N_{\rm in}=N_B$.
			
			\State Choose a trial endpoint $N_{\rm out}<N_B$
			such that $L(N_{\rm out})<c$.
			
			\If{no such physical endpoint exists for $N_s\ge0$}
			\State Set $N_{s,\min}=0$.
			\Else
			
			\While{$|N_{\rm in}-N_{\rm out}|>\mathrm{Tol}$}
			
			\State Set $N_{\rm mid}=\frac{N_{\rm in}+N_{\rm out}}{2}$.
			
			\If{$L(N_{\rm mid})\ge c$}
			\State $N_{\rm in}\gets N_{\rm mid}$.
			\Else
			\State $N_{\rm out}\gets N_{\rm mid}$.
			\EndIf
			
			\EndWhile
			
			\State Set $N_{s,\min}=\frac{N_{\rm in}+N_{\rm out}}{2}$.
			
			\EndIf
			
			\State \Return $N_{s,\max}$ and $N_{s,\min}$.
			
		\end{algorithmic}
		\vspace{5pt}
		\hrule
		\vspace{2pt}
		\hrule
	\end{minipage}
\end{figure}

\section{Quantum Fisher information with mean vector and Wigner covariance matrix}\label{App:QFI}
For an $n$-mode Gaussian state $\rho_x$ depending on an unknown parameter $x$, the quantum Fisher information (QFI) is defined as
\begin{eqnarray}\label{Eq:QFI}
F_x=\text{Tr}[\rho_x L_x^2],
\end{eqnarray} 
where $L_x$ is known as the symmetric logarithmic derivative (SLD) operator satisfying the Lyapunov formula
\begin{eqnarray}
\partial_x \rho_x =\frac{\rho_x L_x+ L_x \rho_x }{2}.
\end{eqnarray} 

Given the mean vector $\bar{\bm{Q}}$ and Wigner covariance matrix $\bm{V}$, the SLD operator can be expressed in terms of the quadrature operators as~\cite{QP,ge}
\begin{eqnarray}\label{eq:SLD}
L_x=\Delta \bm{Q}^T \bm{\Phi}_x \Delta \bm{Q}+ \bm{\xi}_x^T \Delta \mathbf{Q} -\text{Tr}[\bm{\Phi}_x \mathbf{V}],
\end{eqnarray}
with
\begin{eqnarray}
\Delta \mathbf{Q}&=&\mathbf{Q}-\bar{\mathbf{Q}}\;,\nonumber\\
\bm{\xi}_x&=&\bm{V}^{-1} \partial_x \bar{\bm{Q}}\;,
\end{eqnarray}
and the symmetric matrix $\bm{\Phi}_x$ is determined by
\begin{equation}\label{eq:shiyan}
\partial_x\bm{V}=2\bm{V}\bm{\Phi}_x\bm{V}-\frac{1}{2}\bm{\Omega}\bm{\Phi}_x\bm{\Omega}^{T}.
\end{equation}
Inserting Eqs.~(\ref{eq:SLD}) into (\ref{Eq:QFI}), one has
\begin{eqnarray}\label{eq:FQQ}
F_x= \text{Tr}[(\partial_x \bm{V}) \bm{\Phi}_x]+(\partial_x \bar{\mathbf{Q}})^T \bm{V}^{-1} (\partial_x \bar{\mathbf{Q}}),
\end{eqnarray}
which is derived explicitly in Ref.~\cite{QP}.

To obtain a compact form, we vectorize the matrix equation using the map $(B \otimes A) \; \text{vec}[X]=\text{vec}\left[A X B^T\right]$ for the matrices $A$, $B$ and $X$. 
Here, $\operatorname{vec}[X]$ denotes the column-wise vectorization of a matrix $X$, obtained by stacking its columns into a single column vector. For example, the column-wise vectorization for $X=\left(\begin{matrix}x_{11} & x_{12}\\ x_{21} & x_{22} \end{matrix}\right)$ is $\text{vec}[X]=\left(x_{11}, x_{21},x_{12}, x_{22} \right)^T$.
Hence, we vectorize Eq.~(\ref{eq:shiyan}) as
\begin{eqnarray}
&&\text{vec}[\partial_x \bm{V}]=2 \left[\bm{V} \otimes \bm{V} -\frac{1}{4} \bm{\Omega} \otimes \bm{\Omega} \right] \text{vec}[\bm{\Phi}_x],\nonumber\\
&&\rightarrow \text{vec}[\bm{\Phi}_x]=\frac{1}{2}  \left[\bm{V} \otimes \bm{V} -\frac{1}{4} \bm{\Omega} \otimes \bm{\Omega} \right]^{-1} \text{vec}[\partial_x \bm{V}] .
\end{eqnarray}
Then we use the property $\text{Tr}[A^T B]=\left(\text{vec}[A]\right)^T \text{vec}[B]$ obtain $\text{Tr}[(\partial_x \bm{V}) \bm{\Phi}_x]=\left(\text{vec}[\partial_x \bm{V}]\right)^T \text{vec}[\bm{\Phi}_x]$.
Combining these results, Eq.~(\ref{eq:FQQ}) can be rewritten as
\begin{eqnarray}\label{eq:Fx}
F_x&=&\frac{1}{2} \left(\operatorname{vec}[\partial_x\bm V]\right)^T [\bm{V}\otimes\bm{V}-\frac{1}{4}{\bm{\Omega} \otimes \bm{\Omega}}]^{-1} \text{vec}[\partial_x \bm{V}]\nonumber\\
&+&(\partial_x\bar{\bm{Q}})^T \bm{V}^{-1} (\partial_x \bar{\bm{Q}}).
\end{eqnarray}
For the zero-mean Gaussian states considered here, $\bar{\bm Q}=\bm 0$, such that Eq.~(\ref{eq:Fx}) reduces to
\begin{eqnarray}\label{eq:Fxzero}
{\hspace{-0.8cm}}F_x=\frac{1}{2} \left( \operatorname{vec}[\partial_x\bm V]\right)^T [\bm{V}\otimes\bm{V}-\frac{1}{4}{\bm{\Omega} \otimes \bm{\Omega}}]^{-1} \text{vec}[\partial_x \bm{V}]. 
\end{eqnarray}
We next apply Eq.~(\ref{eq:Fxzero}) to the GCS and TMSV probes.

\paragraph{Gaussian-distributed coherent-state probe.}
For Charlie with the GCS probe, from Eq.~(\ref{eq:QVC}) we get $\mathbf{V}_\text{GCS}^C(\eta)= \left( \eta N_s+(1-\eta)N_B+\frac{1}{2 }\right) \mathbf{I}_{2 \times 2}$.
Then we get $\partial_\eta \mathbf{V}_\text{GCS}^C(\eta)=(N_s-N_B)\mathbf{I}_{2 \times 2}$ and $\text{vec}[\partial_\eta \mathbf{V}_\text{GCS}^C(\eta)]=(N_s-N_B)(1,0,0,1)^T$.
Substituting these results into Eq.~(\ref{eq:Fxzero}), we have the single-copy QFI
\begin{eqnarray}
F_{\eta,1}^\text{GCS}=\frac{ (N_s-N_B)^2}{N_1 (N_1+1)},
\end{eqnarray}
for $N_1=\eta N_s+(1-\eta)N_B$.

For a product state $\rho_x^{(M)}=\bigotimes_{m=1}^{M}\rho_x^{(m)}$, the QFI satisfies the additivity relation
\begin{eqnarray}\label{eq:property}
	F_x\!\left[\rho_x^{(M)}\right]=\sum_{m=1}^{M}F_x\!\left[\rho_x^{(m)}\right].
\end{eqnarray}
Hence, 
the $M$-copy QFI writes
\begin{eqnarray}
	F_{\eta}^{\mathrm{GCS}}=M F_{\eta,1}^{\mathrm{GCS}},
\end{eqnarray}
which gives Eq.~(\ref{eq:FQ}) in the main text.

\paragraph{Two-mode squeezed-vacuum probe.}
For Charlie with the TMSV probe,  from Eq.~(\ref{eq:ss}) we have
$\mathbf{V}_\text{TMSV}^C(\eta)=\left(\begin{matrix} B & C\\ C & A \end{matrix}\right) \oplus \left(\begin{matrix} B& -C\\ -C & A \end{matrix}\right)$ with $A=\eta N_s+(1-\eta)N_B+1/2$, $B=N_s+1/2$, $C=\sqrt{\eta N_s (N_s+1)}$. 
Its derivative with respect to $\eta$ writes
\begin{eqnarray}\label{eq:a}
\partial_\eta \mathbf V_{\mathrm{TMSV}}^C=
	\begin{pmatrix}
		0&E\\
		E&K
	\end{pmatrix}
	\oplus
	\begin{pmatrix}
		0&-E\\
		-E&K
	\end{pmatrix},
\end{eqnarray}
for $E=\frac{\sqrt{N_s(N_s+1)}}{2\sqrt{\eta}}$ and $K=N_s-N_B$. The corresponding vectorization is $\text{vec}[\partial_\eta \mathbf V_{\mathrm{TMSV}}^C]=(0,E,0,0,E,K,0,0,0,0,0,-E,0,0,-E,K)^T$. Substituting Eqs.~(\ref{eq:a}) into (\ref{eq:Fxzero}), the single-copy QFI therefore is
\begin{eqnarray}
{\hspace{-0.5cm}}	F_{\eta,1}^{\mathrm{TMSV}}=
	\frac{
		(1-\eta)N_s(N_s+1)
		+
		\eta[(2N_B+1)N_s+N_B]
	}{
		\eta(1-\eta)
		\{1+(1-\eta)[N_s+N_B+2N_sN_B]\}
	}. \nonumber\\
\end{eqnarray}
We next use the additivity property of Eq.~(\ref{eq:property}), the M-copy QFI is obtained as Eq.~(\ref{eq:t}) in the main text.

\section{Classical Fisher information under probe-energy constraints}\label{App:CFI}
We now derive the classical Fisher information (CFI) in two specific measurement schemes for the GCS and TMSV probes, respectively.

\paragraph{Homodyne measurement.}
For Gaussian probes under homodyne detection, the measurement outcome distribution $p(\bm Q|\eta)=\mathrm{Tr}[\rho_\eta \Pi_{\bm Q}]$  is still Gaussian.
Its mean vector and covariance matrix are determined by the corresponding first and second moments of the Gaussian state $\rho_\eta$. One therefore has
\begin{eqnarray}\label{eq:prob}
	p(\bm Q|\eta)=\frac{\exp\!\left[
		-\frac12
		(\bm Q-\bar{\bm Q})^T
		\bm V^{-1}
		(\bm Q-\bar{\bm Q})
		\right]}{\sqrt{(2\pi)^n \det \bm V}},
\end{eqnarray}
where $n$ denotes the dimension of quadrature vector $\bm{Q}$.
Substituting Eqs.~(\ref{eq:prob}) into the definition (\ref{eq:CFI}), the corresponding CFI is~\cite{CFI1,CFI2,Kay1993}
\begin{eqnarray}
	J_\eta=(\partial_\eta \bar{\bm{Q}})^T \bm{V}^{-1} (\partial_\eta \bar{\bm{Q}})+\frac{1}{2}\text{Tr}[(\bm{V}^{-1} \partial_\eta \bm{V})^2].
\end{eqnarray}
For a zero-mean Gaussian distribution ($\bar{\bm Q}=0$ for all $\eta$), this CFI reduces to
\begin{eqnarray}\label{eq:CFInew}
	J_\eta=\frac{1}{2}\text{Tr}[(\bm{V}^{-1} \partial_\eta \bm{V})^2].
\end{eqnarray}
In the following, we investigate the specific probe states.

For Charlie with the GCS probe, the measured quadrature $\bm{Q}=q$.
The corresponding covariance matrix reduces to $V=\eta N_s+(1-\eta)N_B+1/2$.
Substituting this result into Eq.~(\ref{eq:CFInew}), the M-copy CFI $J_{\eta}^\text{GCS,HD}$ is written as Eq.~(\ref{eq:CFI_GCS}) in the main text.
Here we have used the additivity of the CFI for independent measurement outcomes~\cite{Liu2020}.
Under the probe-energy constraints, the maximum value of $J_{\eta}^\text{GCS,HD}$ is obtained at $N_s=N_B-\Delta_\text{max}$  with $\Delta_\text{max}=\frac{4}{1-\eta} \sqrt{\frac{\epsilon N_B(N_B+1)}{M} }$.  In the strong-covertness regime $\epsilon \ll 1$, we obtain the leading order expression for the CFI as shown in Eq.~(\ref{eq:GC}) in the main text.
Comparing Eq.~(\ref{eq:GC}) to the GCS CCQFI of Eq.~(\ref{eq:ep}), the homodyne attainability is given in Eq.~(\ref{eq:R1}) in the main text.

By contrast, for Charlie with the TMSV probe, we consider joint homodyne detection～\cite{JointHD1,JointHD2} of the $q$ quadratures of the idler and return modes, namely $q_I$ and $q_R$. 
The measured quadratures are $\bm{Q}=(q_I,q_R)^T$ with zero mean value. 
The covariance matrix is given by the $q$-quadrature block matrix, i.e., 
$\bm{V}=\begin{pmatrix}
		B & C\\
		C & A
\end{pmatrix}$ with $A=\eta N_s+(1-\eta)N_B+\frac{1}{2},B=N_s+\frac{1}{2},C=\sqrt{\eta N_s(N_s+1)}$.
Substituting this result into Eq.~(\ref{eq:CFInew}), the corresponding $M$-copy CFI is written into
\begin{eqnarray}\label{eq:CFI_TMSV_HD}
J_{\eta}^{\mathrm{TMSV,HD}}
&=&\frac{M(N_s+N_B+2N_s N_B)^2}{8(AB-C^2)^2}\nonumber\\
&+&\frac{MN_s(N_s+1)}{4\eta (AB-C^2)}. 
\end{eqnarray}
Within the probe-energy-constrainted window, the maximum of Eq.~(\ref{eq:CFI_TMSV_HD}) is obtained at $N_s=N_B+\Delta_\text{max}$. 
We therefore further get
\begin{eqnarray}
&&\mathcal{J}_\eta^\text{TMSV,HD}  \simeq  \frac{M S[1+4(1+\eta)S]}{\eta(2N_\Delta-1)^2}\nonumber\\
&&+\sqrt{\epsilon M S} \frac{4(2N_B+1)[1+2(3+\eta)S+8(1-\eta)^2 S^2 ]}{\eta(1-\eta)(2N_\Delta-1)^3} \nonumber\\
&&+\frac{16\epsilon S \Gamma}{\eta(1-\eta)^2 (2N_\Delta-1)^4},
\end{eqnarray}
for $S=N_B(N_B+1)$, $\Gamma=16\eta^3 S^2+16\eta^2S^2+4\eta^2S-16\eta S^2+12\eta S+4\eta-16S^2-8S-1$ and $N_\Delta=1+2(1-\eta)N_B(N_B+1)$.
Comparing to the TMSV CCQFI of Eq.~(\ref{eq:CCQFITMSVexp}), the corresponding homodyne attainability $\mathcal{R}^\text{TMSV,HD}$ can be obtained, whose expression is cumbersome and is numerically computed.

\paragraph{Photon-counting measurement.}
In photon-counting measurements, photon-number-resolving (PNR) detectors are used to obtain discrete photon-number outcomes.
Therefore, the Gaussian-outcome CFI formula in Eq.~(\ref{eq:CFInew}) is no longer applicable, the relevant CFI is determined from the discrete photon-number statistics.
For joint photon counting of $K$ modes, the corresponding CFI is
\begin{eqnarray}\label{eq:CFIdis}
	J_\eta
	=
	\sum_{\mathbf{n}\in\mathbb{N}_0^K}
	\frac{
		[\partial_\eta p(\mathbf{n}|\eta)]^2
	}{
		p(\mathbf{n}|\eta)
	},
\end{eqnarray}
where $\mathbb{N}_0^K$ denotes the set of all $K$-dimensional vectors whose entries are nonnegative integers. 
Specifically, $\mathbf{n}=(n_1,n_2,\ldots,n_K)$ denotes the photon-number vector of the measured modes.

For Charlie with the GCS probe, the returned signal is in a thermal state with single mode $\bm{n}=(n)$. 
Its photon-number distribution is
\begin{eqnarray}
p^{\rm GCS,PC}(n|\eta)=\frac{N_1^{n}}{(N_1+1)^{{n}+1}},
\end{eqnarray}
for $N_1=\eta N_s+(1-\eta)N_B$.
Substituting this distribution into Eq.~(\ref{eq:CFIdis}), the $M$-copy CFI is given by Eq.~(\ref{eq:GCSCFI}) in the main text.
This exactly coincides with the corresponding GCS QFI in Eq.~(\ref{eq:FQ}).
Hence, maximizing both quantities over the same covertness-induced probe-energy window gives the same result, as shown in Eq.~(\ref{eq:GCSHD}) of the main text.

In contrast, for the TMSV probe, we consider joint photon-number-resolving detection of the idler and return modes.
The joint photon-number outcome is $\bm{n}=(n,n')$, where Charlie detects $n$ photons in the idler mode and $n'$ photons in the return mode.
Its joint photon-number distribution is
\begin{equation}\label{eq:probnew}
	p^{\rm TMSV,PC}(n,n'|\eta)=p_I(n)\,p_R(n'|n,\eta),
\end{equation}
where $p_I(n)=\langle n|\text{Tr}_s[|\psi_\text{TMSV}\rangle \langle \psi_\text{TMSV}|]|n\rangle=\frac{N_s^n}{(N_s+1)^{n+1}}$ is the photon-number distribution of the idler mode, and $p_R(n'|n,\eta)$ denotes the conditional probability of detecting $n'$ photons in the return mode given $n$ photons in the idler mode.

The conditional probability can be further expressed as a sum over products of a binomial distribution $\mathfrak{p}(k|n)$ and a negative-binomial distribution $\mathfrak{p}(n'|k)$,
\begin{eqnarray}\label{eq:Dform}
	p_R(n'|n,\eta)
	&=&
	\sum_{k=0}^{\min\{n,n'\}}
	\mathfrak{p}(k|n)\,
	\mathfrak{p}(n'|k)
	\nonumber\\
	&=&
	\sum_{k=0}^{\min\{n,n'\}}
	\binom{n}{k}
	x^k(1-x)^{n-k}
	\nonumber\\
	&&\times
	(1-y)
	\binom{n'}{k}
	(1-y)^k
	y^{\,n'-k},
\end{eqnarray}
for $x=\frac{\eta}{1+(1-\eta)N_B},y=\frac{(1-\eta)N_B}{1+(1-\eta)N_B}$. 
In the noiseless limit $N_B=0$, one has $x=\eta$ and $y=0$.
Thus, $x$ can be interpreted as an effective reflectivity parameter dressed by the thermal background, whereas $y$ characterizes the thermal-noise contribution at the return port.

Substituting Eq.~(\ref{eq:probnew}) into Eq.~(\ref{eq:CFIdis}), and noting that $p_I(n)$ is independent of $\eta$, the $M$-copy CFI becomes
\begin{eqnarray}\label{eq:CFI_TMSV_PC}
	J_{\eta}^{\rm TMSV,PC}
	=
	M
	\sum_{n=0}^{\infty}
	p_I(n)
	\sum_{n'=0}^{\infty}
	\frac{
		\left[
		\partial_\eta p_R(n'|n,\eta)
		\right]^2
	}{
		p_R(n'|n,\eta)
	}.
\end{eqnarray}
For the numerical evaluation of Eq.~(\ref{eq:CFI_TMSV_PC}), the photon-number sums are truncated at $n_{\max}=20$ and $n'_{\max}=30$.
Terms with $p_R(n'|n,\eta)<10^{-14}$ are discarded to avoid numerical instability associated with division by extremely small probabilities.
\begin{figure}[!h]
	\centering
	\includegraphics[width=0.5\textwidth]{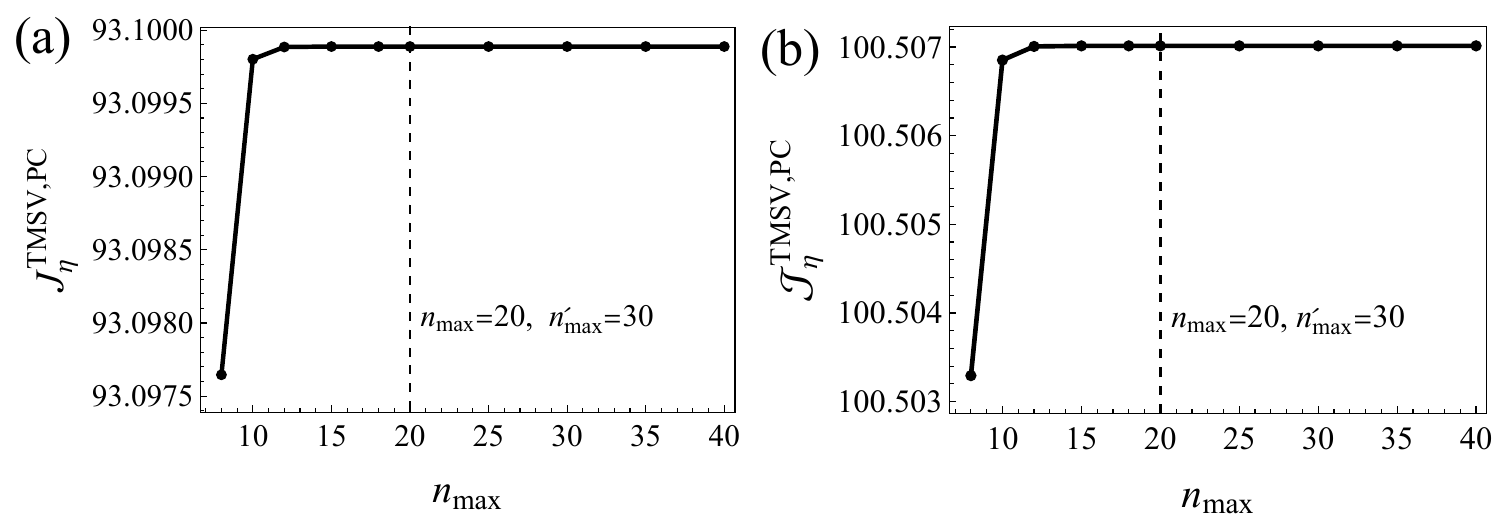}
	\caption{
		Numerical convergence of the TMSV classical Fisher information under photon-counting measurement.
		(a) CFI $J_{\eta}^{\rm TMSV,PC}$ at the perfect-covertness point $N_s=N_B=0.2$, and
		(b) covertness-constrained CFI $\mathcal{J}_{\eta}^{\rm TMSV,PC}$ evaluated at $N_s=N_{s,\max}(\epsilon)$, as functions of the photon-number cutoff $n_{\max}$.
		The cutoff $n'_{\max}$ is increased simultaneously with $n_{\max}$, with $n'_{\max}\simeq1.5n_{\max}$.
		The dashed line indicates the truncation $(n_{\max},n'_{\max})=(20,30)$ adopted in the numerical calculations.
		The common parameters are $\eta=0.01$, $N_B=0.2$, $M=100$ and in panel (b) $\epsilon=0.005$.
	}
	\label{Fig.con}
\end{figure}

In addition, as shown in Fig.~\ref{Fig.con} (a), the numerical convergence with respect to the photon-number truncation is verified by simultaneously increasing $n_{\max}$ and $n'_\text{max}$ with $n'_\text{max}\simeq 1.5 n_{\max}$. 
The CFI $J_{\eta}^{\rm TMSV,PC}$ rapidly converges to a stable value, and our choice of $n_{\max}=20, n'_{\max}=30$ is sufficient.
Under the restricted probe energy, the corresponding covertness-constrained CFI $\mathcal{J}_\eta^\text{TMSV,PC}$ involves the same infinite double sum and is evaluated using the same photon-number truncation.
Its convergence is shown in Fig.~\ref{Fig.con} (b), where it likewise approaches a stable value.
Its asymptotic scaling is given by Eq.~(\ref{eq:QFIscaling}) in the main text. 
The corresponding PC attainability $\mathcal{R}^\text{TMSV,PC}$ is therefore evaluated numerically.
For the parameter ranges considered in this work, these truncations give converged results.

\bibliography{Maintext}

\end{document}